\documentclass[twocolumn]{aastex62}
\usepackage{times,graphics}
\usepackage{hyperref}

\usepackage{graphicx}
\usepackage{dcolumn}
\usepackage{bm}
\usepackage{natbib}
\usepackage{pstricks}
\usepackage{epsfig}
\usepackage{epstopdf}
\usepackage{multirow}
\usepackage{amsmath}
\usepackage{lipsum}

\graphicspath{{figs/}}

\begin{document}

\title{ \bf \Large {Multiple transitions in vacuum dark energy and $H_0$ tension}}

	\author{Hossein Moshafi}
	\email{moshafi@ipm.ir}
	\affiliation{School of Astronomy, Institute for Research in Fundamental Sciences (IPM) \\ P.~O.~Box 19395-5531, Tehran, Iran}
	
	\author{Hassan Firouzjahi}
	\email{firouz@ipm.ir}
	\affiliation{School of Astronomy, Institute for Research in Fundamental Sciences (IPM) \\ P.~O.~Box 19395-5531, Tehran, Iran}
	
	\author{Alireza Talebian}
	\email{talebian@ipm.ir}
	\affiliation{School of Astronomy, Institute for Research in Fundamental Sciences (IPM) \\ P.~O.~Box 19395-5531, Tehran, Iran}

	\date{\today}

\begin{abstract}
We study the effects of multiple transitions in the vacuum dark energy density on the $H_0$ tension problem. We consider a phenomenological model in which the vacuum energy density undergoes multiple transitions in the early as well as the late universe and compare the model's predictions using the three sets of data from CMB+BAO+SN.  The transient dark energy can be either positive (dS-like) or negative (AdS-like). We conclude that a transient late-time AdS-type vacuum energy typically yields the higher value of $H_0$ which can alleviate the $H_0$ tension. In addition, to obtain a value of $H_0$ comparable to the value obtained from the local cosmological measurements  the spectral index $n_s$ moves towards its Harrison-Zel'dovich scale invariant value. 
\end{abstract}

\section{Introduction}
	
The disagreement between the independent measurements of the Hubble 
constant, based on the early universe with the $\Lambda{\rm CDM}$ model and that measured by direct
observations of the local universe without assuming the $\Lambda{\rm CDM}$ model,  is dubbed as  the Hubble tension~\citep{DiValentino:2021izs, Perivolaropoulos:2021jda, Verde:2019ivm, Riess:2019qba, DiValentino:2020zio, Schoneberg:2021qvd, Dainotti:2021pqg, Dainotti:2022bzg}. More precisely, the reported value of the Hubble parameter from the Planck satellite is $H_0 = 67.4 \pm 0.5 ~{\rm Km/s/Mpc}$~\citep{Planck:2018vyg} while the latest SH0ES Team~\citep{Riess:2021jrx} has reported  $H_0 = 73.30 \pm 1.04 ~{\rm Km/s/Mpc}$. This tension is an important open problem in cosmology.  
 Ignoring the possible systematics origin of the tension~\citep{Freedman:2019jwv,Efstathiou:2020wxn,Mortsell:2021nzg}, it is natural to expect that either a modification to the cosmological model is required or there is a new physics behind the tension~\citep{Mortsell:2018mfj,Guo:2018ans,Vagnozzi:2019ezj,DiValentino:2016hlg,Knox:2019rjx,Schoneberg:2021qvd}.

Qualitatively speaking, there are two classes of models attempting to resolve the Hubble tension by introducing new physics. The first class of models are based on the  modifications in dark components at late times (i.e. , lower redshifts), e.g. , by introducing a dynamical dark energy
that alter the Hubble expansion. 
In the second class of models, the new physics aims to reduce the sound horizon by $\sim 7\%$~\citep{Bernal:2016gxb,Lemos:2018smw,Knox:2019rjx} while keeping the  Baryon Acoustic Oscillation (BAO) and the uncalibrated SNeIa data to be consistent with Planck measurements. The locations of the acoustic peaks in the CMB observations is one of the most precisely measured quantity in cosmology. 
With an accuracy of $0.03\%$ the Planck satellite  \citep{Planck:2018vyg} has determined the angular size of the sound horizon at recombination,
$\theta_\star \equiv r_{\star}/D_{\star}$, in which the sound horizon $r_{\star}$ is the comoving distance a sound wave could travel from the beginning of the universe to the time of recombination, 
and $D_{\star}$ is the comoving integrated distance from now to 
the epoch of recombination. $D_{\star}$ depends only on two parameters $\Omega_m$ (the fractional matter energy density today) and the present value of Hubble expansion rate, $H_0$. Thus, given $r_{\star}$ and an estimation of $\Omega_m$, one can infer $H_0$ from the measurement of $\theta_\star$. Using the Planck best fit values of $\Omega_m$ and $r_\star$ in the context of the $\Lambda$CDM model \citep{Planck:2018vyg}, $H_0$ is found to be significantly lower than the more direct local measurements. If the value of the Hubble constant from SH0ES is considered, it would yield a much larger value of $\theta_\star$ unless either $r_\star$ and/or $D_\star$ were modified to preserve the observed CMB acoustic peak positions.

Pre-recombination early dark energy (EDE)~\citep{Karwal:2016vyq,Poulin:2018cxd,Kaloper:2019lpl,Agrawal:2019lmo,Lin:2019qug,Smith:2019ihp,Niedermann:2019olb,Sakstein:2019fmf,Ye:2020btb,Gogoi:2020qif,Braglia:2020bym,Lin:2020jcb,Seto:2021xua,Nojiri:2021dze,Karwal:2021vpk} is one of the best-studied scenario as a solution to the Hubble tension. In this scenario, a dark energy-like component is introduced. The energy injection before recombination boosts the Hubble expansion rate (by reducing the sound horizon). The EDE then decays rapidly in order not to spoil other observations. Various scenarios have been proposed for both the early energy injection~\citep{Poulin:2018cxd,Kaloper:2019lpl,Agrawal:2019lmo,Lin:2019qug,Smith:2019ihp,Niedermann:2019olb,Sakstein:2019fmf,Ye:2020btb,Gogoi:2020qif,Braglia:2020bym,Lin:2020jcb,Seto:2021xua,Nojiri:2021dze,Karwal:2021vpk,Zumalacarregui:2020cjh,Ballesteros:2020sik,Braglia:2020auw} and the decaying processes~\citep{Poulin:2018cxd,Smith:2019ihp,Ye:2020btb}.

 EDE models have important effects on  primordial scalar perturbations and on Large-Scale Structure (LSS) physics. It has been found that the EDE models require a re-interpretation of the available data, resulting in higher values of $n_s$, all the way up to a scale-invariant Harrison-Zeldovich
spectrum of primordial scalar perturbation, i.e. $n_s= 1$ ~\citep{DiValentino:2018zjj,Ye:2021nej,Jiang:2022uyg}. The second effect of EDE models appears when one considers the galaxy clustering data~\citep{Krishnan:2020obg,Hill:2020osr,Ivanov:2020ril,DAmico:2020ods}. Although  
a large EDE fractions $f_{\rm EDE}$ are not ruled out by these datasets~\citep{Murgia:2020ryi,Smith:2020rxx,Herold:2021ksg,Gomez-Valent:2022hkb}, increasing $f_{\rm EDE}$ will increase the clustering amplitude $\sigma_8$ and the related value of $S_8$~\citep{DiValentino:2018gcu,Nunes:2021ipq}. In other word, the models with a large fractions $f_{\rm EDE}$ worsen the well-known ``$S_8$ discrepancy''. Most of EDE models proposed as a solution to the Hubble tension leads to an  increase of clustering amplitude ($S_8$) that worsen the fit to galaxy clustering data. Ref.~\citep{Reeves:2022aoi} argued that freeing the total neutrino mass $M_\nu$ can suppress small-scale power and then improve EDE's fit to galaxy clustering data.

Currently, the most precise large-scale CMB observation is Planck data, which alone seems not to favour axion-like EDE models~\citep{Hill:2020osr}. However, the TT power spectrum of Planck data, especially in small scales, needs more considerations. Some inconsistencies between the ${\ell_\mathrm{TT}}<1000$ and ${\ell_\mathrm{TT}}>1000$ part of Planck's TT power spectrum have been reported 
in Refs.~\citep{Addison:2015wyg,Planck:2016tof}. 
Moreover, the amplitude of CMB gravitational lensing in Planck data is not consistent with what we expect in $\Lambda$CDM model. The smoothing effect of gravitational lensing on acoustic peaks of the
CMB power spectrum exceeds that expected in $\Lambda$CDM model
\citep{Addison:2015wyg,Motloch:2019gux}. However, small scale ground-based CMB
observations such as ACT and SPT, providing precise
measurements on small scale power spectrum, have not found
this over-smoothing effect~\citep{SPT:2017jdf,ACT:2020gnv,SPT-3G:2021eoc}.
Recently, combined analysis of Planck ($\ell_\mathrm{TT} \lesssim 1000$) with ACT and SPT data
for EDE models has also been performed, such as CMB+SPTpol
\citep{Chudaykin:2020acu,Chudaykin:2020igl,Jiang:2021bab},
CMB+ACT DR4 \citep{Hill:2021yec,Poulin:2021bjr} and CMB+ACT
DR4+SPT-3G \citep{LaPosta:2021pgm,Smith:2022hwi,Jiang:2022uyg}. 
Also for CMB+LSS data see~\citep{Hill:2020osr,Ivanov:2020ril,DAmico:2020ods,Ye:2021iwa}.

In this paper, we study a scenario with multiple transient 
phases of dark energy, both before and after the surface of last scattering,
yielding to a higher value of the current Hubble expansion rate compared to what is inferred from the $\Lambda {\rm CDM}$ model.  This is a phenomenological model motivated from \citep{Firouzjahi:2022xxb} where the quantum vacuum zero point energy in connection to cosmological constant problem and the origin of dark energy were studied. Alternatively,  the current setup may be viewed as an independent phenomenological mechanism with some similarities to  the EDE proposal.  For other works involving phase transitions in dark energy  and their impacts on $H_0$ tension see also 
\citep{Banihashemi:2018oxo, Farhang:2020sij, Khosravi:2021csn, Moshafi:2020rkq}.

\section{The Model}
\label{sec:model}

We consider the model consisting of some new energy density source $\rho_{_X}$ arising from the quantum zero point energy of the fields with the mass $m$. Depending   on the type of the quantum field (boson or fermion) and the energy scale of interest $\rho_{_X}$ can be either positive (dS-like) 
or negative (AdS-like)  \citep{Martin:2012bt, Firouzjahi:2022xxb, Firouzjahi:2022vij}.  
 For the sake of simplicity, here first we consider the case of a single quantum field  while the procedure for the multiple field is straightforward thanks to our simplifying  assumption that the vacuum energies of the free fields do not exchange with each other.

\begin{figure}[h]
	\centering
	\includegraphics[width=\linewidth]{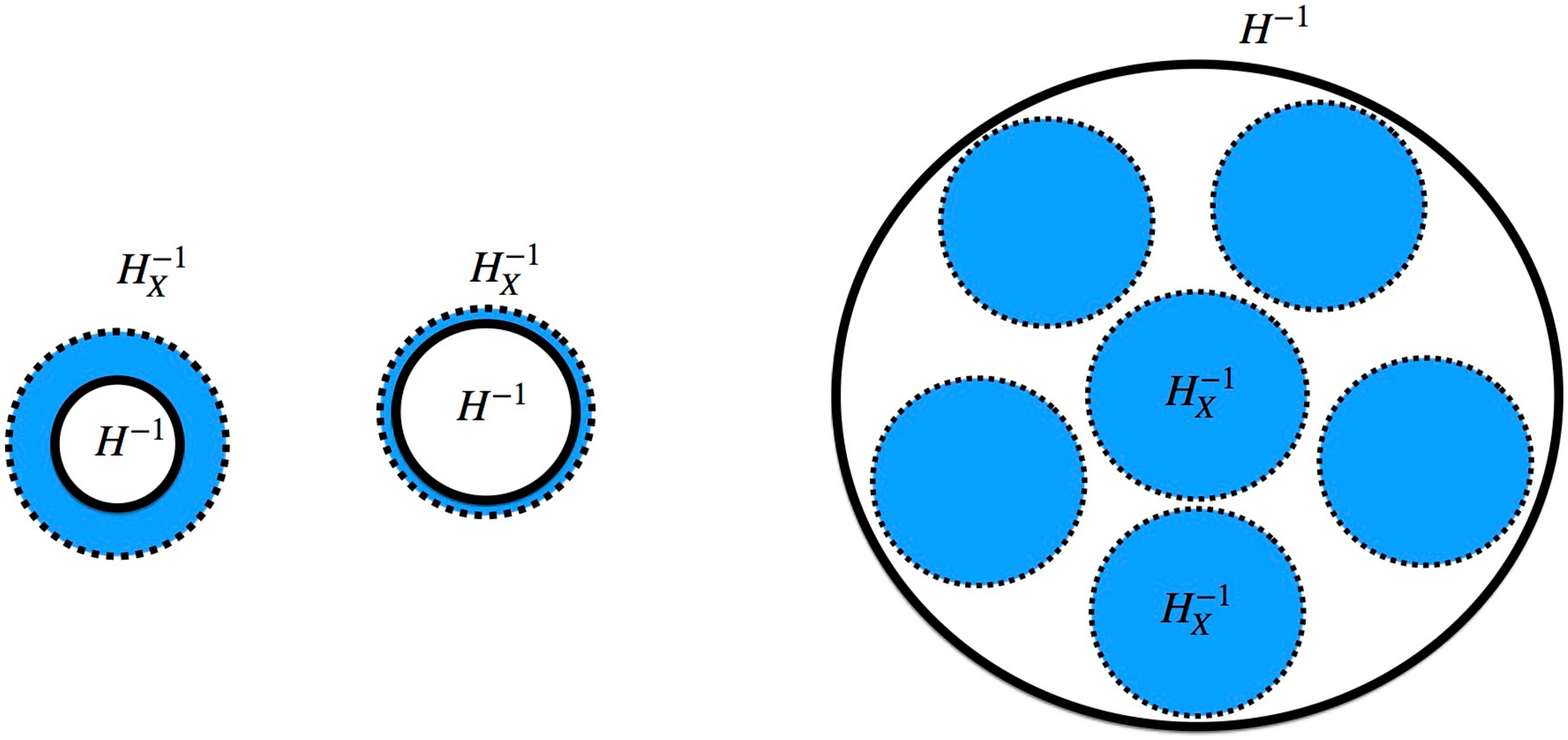}
	\caption{A schematic view of the evolution of dark energy $\rho_{_X}$ compared to total energy energy density $\rho$. Left: at early times $\rho_{_X} \ll \rho$ and $H_X^{-1} \gg H^{-1}$ so the FLRW horizon is within one patch of dS horizon associated to dark energy. Middle:   around the time of transition $\rho_{_X} \sim \rho $ and $H_X^{-1} \sim H^{-1}$. This is the time when the effects of dark energy become important. Right: long after transition, many patches of dS horizon associated to $\rho_{_X}$ enter the FLRW horizon and the effects of $\rho_{_X}$ are represented by a phenomenological fluid with the equation of state $w$. }
	\label{fig:H-1}
\end{figure}


Associated to the energy density $\rho_{_X}$, we can define a horizon radius, denoted  by $H_{_X}^{-1}$, in which $3 M_P^2 H_{_X}^2 = |\rho_{_X}|$ in which $M_P$ is the reduced Planck mass.  Let us also denote $H$ as the Hubble rate of the observable FLRW universe with $3M_P^2 H^2 = \rho$ in which $\rho$ is the total energy density.  
Depending on the mass scale $m$, the contribution of $\rho_{_X}$ can be viewed as a  source of dark energy in  $\rho$. At the early stage in cosmic expansion history the zero point energy of the field appears as a small constant term in $\rho$.
More specifically, $\rho_{_X} = c \, m^4$ with $c$ is a constant. As  illustrated in left panel of Fig.~\ref{fig:H-1} in this regime  $H^{-1} \ll H_{_X}^{-1}$ so the  FLRW horizon (at that time) is within one patch of the vacuum energy.  As time proceeds and $\rho$ decreases further, we can indicate a specific scale factor $a_{\rm c}$ (or redshift $z_{\rm c})$  when $\rho(a_{\rm c}) \sim \rho_{_X} \sim m^4$. At this stage in cosmic epoch $ H_{_X}^{-1} \sim H^{-1}$, as in the middle panel of Fig.~\ref{fig:H-1}, and 
$\rho_{_X}$ can be relevant as the source of dark energy at that time.  
 After this phase, $\rho$ falls off rapidly and very soon we have $H^{-1} \gg H_{_X}^{-1}$, as illustrated in right panel of Fig.~\ref{fig:H-1}. 
It is argued in \citep{Firouzjahi:2022xxb} that  as many patches of vacuum horizon  
with the size $H_{_X}^{-1}$ enters the FLRW horizon the patches of zero point energy develops inhomogeneities $ \frac{\delta \rho_{_X}}{\rho_{_X}} > 1 $. The subsequent  evolution of $\rho_{_X}$ on sub Hubble scale is left as an open problem in \citep{Firouzjahi:2022xxb}. The time of the transition in dark energy
is determined by the mass of the quantum field, happening at the temperature
$T_c\sim m$. 
For example, for the simplest model in Standard Model (SM) 
with three massive neutrinos, we can have three transitions in dark energy 
in which the first transition  may happen shortly after the time of CMB decoupling while the other two transitions may happen much later in cosmic expansion history. In addition, the current stage of dark energy may be associated with the zero point energy of the lightest neutrino field with the mass $m_{\nu} \sim 10^{-2} \mathrm{eV}$. 

Motivated by the above discussions,  we picture the effects of $\rho_{_X}$ after  the time when $a>a_{\rm c}$ by a phenomenological dark fluid with the equation of state $w$. In this phenomenological view,  $\rho_{_X}$ is constant at early universe ($a \ll a_{\rm c}$) and dilutes as $a^{-3(1+w)}$ at late time ($a \gg a_{\rm c}$).  The requirement that  the energy density of the dark fluid does not overclose the universe too early, we expect that $\rho_{_X}$ to fall off faster than radiation, so we impose $\frac{1}{3} < w \leq 1$. In addition, we keep the position of the transition to be free. In practice, motivated by EDE proposal, we take the position of the first step 
(first transition) to be after the time of matter radiation equality and before the CMB decoupling around the redshift $z_{\rm c}\sim 2000- 3000$ while the follow up transitions can happen much later in cosmic expansion history.   Within the model of \citep{Firouzjahi:2022xxb} this corresponds to a field with the mass $m \sim \mathrm{eV}$. As a field with this mass is not within the spectrum of SM, one needs a new physics beyond SM. For example, this may come from an eV-scale sterile neutrino e.g. from short-baseline anomalies but this is tightly constrained, see for example \citep{Hagstotz:2020ukm}. 

The simplest model satisfying our requirements may be realized by the following ansatz, 
\begin{align}
	\rho_{_X}(a) \propto \Big[1+\big(\dfrac{a}{a_{\rm c}}\big)^{3(1+w)}\Big]^{-1} \,.
	\label{rhoX}
\end{align}
At early stages, $a {\ll} a_{\rm c}$ in which $a_{\rm c}$ is the transition scale factor, $\rho_{_X}(a)$ is nearly constant but small 
while long after the transition it falls off like $a^{-3(1+w)}$.  However, to control the sharpness of dark energy transition we introduce a transfer function ${\cal T}(b,a-a_{\rm c})$ in which  $b$ is a positive constant parameter measuring the sharpness of transition.  In our analysis we consider the following transfer function
\begin{align}
	{\cal T}(b,a-a_{\rm c}) = \dfrac{1}{2}\Big[1+\tanh\big(b\big(a-a_{\rm c}\big)\big)\Big] \,.
\end{align}
For $b |a-a_{ \rm c} |\gg 1$, the transfer function ${\cal T}$ behaves like the Heaviside function $\Theta(a-a_{\rm c})$ indicating a sharp transition 
while for $b |a-a_{\rm c}| \ll 1$, ${\cal T}$ describes a mild transition. 

With the  inclusion of the above transfer function,  Eq. \eqref{rhoX} is modified as
\begin{align}
	\rho_{_X}(a) = \rho_{_X, {\rm c}} \Big[{\cal T}(b,a_{\rm c}-a)+{\cal T}(b,a-a_{\rm c})\big(\dfrac{a}{a_{\rm c}}\big)^{3(1+w)}\Big]^{-1} \, ,
	\label{rhoXm}
\end{align}
where $\rho_{_X, {\rm c}} = \rho_{_X}(a=a_{\rm c})$. At the transition scale factor $a_{\rm c}$, we introduce the fraction of dark energy density $f_{_X}$ associated to the field (or dark fluid)  as
\begin{align}
	f_{_X} \equiv \dfrac{\rho_{_X, {\rm c}}}{\rho(a_{\rm c})} \, ,
\end{align}
where $\rho(a)$ is the total energy density in the Friedmann equation, $3 M_P^2 H^2 = \rho(a)$, and is given by
\begin{align}
\label{rho-a}
	 \rho(a) &= \rho_{\rm m}~a^{-3} + \rho_{\rm r}~a^{-4}+\rho_{\Lambda}+\rho_{_X}(a) \,,
\end{align}
in which $\rho_{\rm m}, \rho_{\rm r}$ and $\rho_\Lambda$ respectively are the present value ($a=1$) of matter, radiation and the dark energy density of the universe. At the transition scale $a_{\rm c}$ we find
\begin{align}
	\rho(a_{\rm c}) &= \dfrac{\rho_{\rm m} \, a_{\rm c}^{-3} + \rho_{\rm r} \, a_{\rm c}^{-4}+\rho_{\Lambda}}{1-f_{_X}} \, ,
\end{align}
and then
\begin{align}
	\rho_{_X, {\rm c}} =
	\dfrac{f_{_X}}{1-f_{_X}} \left( \rho_{\rm m} \, a_{\rm c}^{-3} + \rho_{\rm r} \, a_{\rm c}^{-4}+\rho_{\Lambda} \right) \, .
\end{align}
Note that for dS-like (AdS-like) dark energy $f_{_X} >0 ~ (f_{_X}<0)$.    

Plugging the above value in Eq. (\ref{rho-a}), the total energy density of our model is given by
\begin{eqnarray}
\label{rho-a-final1}
\rho(a) = \rho_{\rm m} \, a^{-3} + \rho_{\rm r} \, a^{-4}+\rho_{\Lambda} + \nonumber \\
\dfrac{\dfrac{f_{_X}}{1-f_{_X}}\Big( \rho_{\rm m} \, a_{\rm c}^{-3} + \rho_{\rm r} \, a_{\rm c}^{-4}+\rho_{\Lambda} \Big) }{{\cal T}(b,a_{\rm c}-a)+{\cal T}(b,a-a_{\rm c})\left(\dfrac{a}{a_{\rm c}}\right)^{3(1+w)}} \, .
\end{eqnarray}

The fraction of the current energy density $\Omega_i$ for each component can  easily be calculated. At present, $3 M_P^2 H_0^2 = \rho_0$, we have
\begin{align}
	\rho_0 = \rho_{\rm m} + \rho_{\rm r}+\rho_{\Lambda}+\rho_{_X,0} \,,
\end{align}
in which $\rho_{_X,0} = \rho_{_X}(a=1)$. Correspondingly, for each component in cosmic fluid we have 
\begin{align}
	\Omega_{\rm m} = \dfrac{\rho_{\rm m}}{\rho_0} {\,,}
	\hspace{.5cm}
	\Omega_{\rm r} = \dfrac{\rho_{\rm r}}{\rho_0} {\,,}
	\hspace{.5cm}
	\Omega_\Lambda = \dfrac{\rho_{\Lambda}}{\rho_0} {\,,}
	\hspace{.5cm}
	\Omega_{_X} = \dfrac{\rho_{_X,0}}{\rho_0} {\,.}
\end{align}

As a rough estimation of $\Omega_{_X}$, using  the simple form of $\rho_{_X}$ from Eq. \eqref{rhoX}, the fraction of dark energy density is obtained to be
\begin{align}
\Omega_{_X} = \frac{2 f_{_X}}{1-f_{_X}} \, \frac{a_{\rm c}^{3 w} }{1+ a_{\rm c}^{3 w}} \Big[ \Omega_{m} +  \Omega_{r} \, a_{\rm c}^{-1} +
\Omega_{\Lambda} a_{\rm c}^{3} \Big] \, .
\end{align}
For the early time when $a_{\rm c} \ll 1$ we approximately obtain, 
\begin{align}
\Omega_{_X} \simeq
\frac{2 f_{_X}}{1-f_{_X}} \,  {a_{\rm c}^{3 w} } \Big[ \Omega_{m} +  \Omega_{r} \, a_{\rm c}^{-1}  \Big] \, .
\end{align}
If we further assume that $a_{\rm c} \ll a_{\rm {eq}}$, i.e. the transition happens after
the matter radiation equality, then the above relation simplifies further to
\begin{align}
\Omega_{_X} \simeq
\frac{2 f_{_X}}{1-f_{_X}} \, {a_{\rm c}^{3 w} } \Omega_{m} \, .
\end{align}
Note that the parameter $\Omega_{_X}$ is what defined in \citep{Karwal:2016vyq}
 as $\Omega_{ee}$. For $a_{\rm c} \lesssim 10^{-3}$ and $f \sim 0.1$, with $w=1$, we typically have $\Omega_{_X} \sim 10^{-10}$ or so. For $a_{\rm c} \sim 10^{-1}$ and $w=1$, we obtain $\Omega_{_X} \sim 10^{-5}$. So in general for $w >\frac{1}{3}$,  $\Omega_{_X}$ is  smaller than $\Omega_r \sim 10^{-4}$.

For the $N$-field configuration, yielding to $N$ transitions in dark energy at 
$a_{{\rm c},i}$ with $i=1,2,.., N$, 
one can extend \eqref{rho-a-final1} to the following form
\begin{eqnarray}
\rho(a) = \rho_{\rm m,0} \, a^{-3} + \rho_{\rm r,0} \, a^{-4}+\rho_{\Lambda}+ \nonumber \\
\sum_{i=1}^{N}\dfrac{\dfrac{f_i}{1-f_i} \left( \rho_{\rm m} \, a_{{\rm c},i}^{-3} + \rho_{\rm r} \, a_{{\rm c},i}^{-4}+\rho_{\Lambda} \right)}{{\cal T}(b_i,a_{{\rm c},i}-a)+{\cal T}(b_i,a-a_{{\rm c},i})\left(\frac{a}{a_{{\rm c},i}}\right)^{3(1+w_i)}}
\end{eqnarray}

In the following analysis, we also parameterize $b_i$ as
\begin{align}
b_i \equiv \dfrac{10^{n_i}}{a_{{\rm c},i}} \, .
\end{align}
For a sharp phase transition in dark energy with $b_i a_{{\rm c},i} \gg 1$, we have 
$n_i >0$ while for a mild transition $n_i <0$.

In summary,  the new parameters of the model are $\{f_i, a_{{\rm c},i}, n_i, w_i \}$ for each component $i=1,2,...N$ of dark fluid.   $f_i$ measures the fraction of dark energy at the time of transition $a_{{\rm c},i}$ (or $z_{{\rm c},i}$), $n_i$ measures the sharpness of the transition and $w_i$ represents the equation of state of the dark fluid after the transition $a(t) > a_{{\rm c},i}$. 

Our model has some similarities to EDE setup in which  we consider an early stage of dark energy. However, we include the new parameter $n_i$ to control how sharp the transition has happened. Moreover, with our theoretical motivations in mind, 
we allow for multiple transitions in dark energy during the Universe evolution, $N \ge 1$. In addition,   the contributions of the subsequent  dark fluids $(1< i \le N)$ can be either  dS ($f_i >0$) or AdS ($f_i<0$). For earlier works concerning AdS vacua 
and $H_0$ tension or AdS-EDE see \cite{Calderon:2020hoc, Ruchika:2020avj,  Dutta:2018vmq, Visinelli:2019qqu, Akarsu:2019hmw, Akarsu:2021fol, Ye:2020btb, Jiang:2021bab, Ye:2020oix, Sen:2021wld}.

There is an important comment in order. In the current analysis of studying the effects of multiple transitions  in dark energy,  we concentrate on the background evolution and do not study perturbations. In our setup the physical mechanism behind the dark energy transitions and the transfer of energy to thermal bath are already complicated phenomena at the background level and a full treatment of perturbations analysis is beyond the scope of the current analysis.  
While studying the background dynamics can shed some light on the $H_0$ tension problem, but it is not fully consistent. This is an important limitation in our current analysis which should be improved in future studies considering the full dynamics of the background and the perturbations.

	
\section{Observational data and Statistical Methodology}
\label{sec:obs}

In this section, we begin with a brief description of the main cosmological data sets used in this work. In all of our analysis we consider a combination of three types of data: ``CMB+BAO+SN''.

\begin{enumerate}

\item {\bf CMB}: We use the latest most precise large-scale cosmic microwave background (CMB) temperature and polarization angular power spectra from the final release of  ``Planck 2018'' \texttt{plikTTTEEE+lowl+lowE} ~\citep{Planck:2018vyg, Planck:2018lbu, Planck:2019nip}. We use full power spectrum and don't split into high-$\ell$ and low-$\ell$ parts. We mention all of Planck data (including temperature and polarization) by ``CMB''.

\item {\bf BAO}: We also take into account the various measurements of the Baryon Acoustic Oscillations (BAO) from different galaxy surveys ~\cite{Planck:2018vyg}, i.e.  6dFGS~\cite{Beutler:2011hx}, SDSS-MGS~\cite{Ross:2014qpa}, and BOSS DR12~\cite{BOSS:2016wmc}.

\item {\bf SN}: We  include the measurements of  the $1048$ Supernovae Type Ia luminosity distance in the red-shift interval $z \in [0.01, 2.3]$, from the Pantheon sample~\citep{Pan-STARRS1:2017jku}. We show this catalog of SuperNovae by ``SN''.

\end{enumerate}
To analyze the data and extract the constraints on the cosmological parameters,  we have modified the well known cosmological MCMC package \texttt{CosmoMC} ~\citep{Lewis:2002ah,Lewis:2013hha}, which is publicly available\footnote{http://cosmologist.info/cosmomc}. 
This package is equipped with a convergence diagnostic based on the Gelman and Rubin statistic ~\citep{Gelman:1992zz}, assuming $R-1 < 0.1$, and implements an efficient sampling of the posterior distribution using the fast/slow parameter decorrelations~ \citep{Lewis:2013hha}. 

Parameter space for $\Lambda$CDM model is:
\begin{eqnarray}
\mathcal{P}_0 \equiv\Bigl\{\Omega_{b}h^2, \Omega_{c}h^2, 100\theta_{\rm MC},
  \tau, n_{s}, \ln[10^{10}A_{s}] \Bigr\}~,
\label{eq:LCDM}
\end{eqnarray}
where $\tau$ is the reionization optical depth, $n_s$ is the scalar
spectral index, $A_{s}$ is the amplitude  of the
scalar primordial power spectrum, and the $\theta_{\rm MC}$ parameter is an approximation of $\theta_*$.

The set of free parameters describing the One-Step class of models (i.e. one transition in dark energy, $N=1$) 
is given by
\begin{equation}
\mathcal{P}_1 \equiv\Bigl\{a_1, f_1, w_1, n_1, \Omega_{b}h^2, \Omega_{c}h^2, 100\theta_{\rm MC},
  \tau, n_{s}, \ln[10^{10}A_{s}] \Bigr\}~  .
\label{eq:one-step-parameters}
\end{equation}
Correspondingly, the parameter space of Two-Step class of models ($N=2$) is given by
\begin{eqnarray}
\mathcal{P}_2 \equiv\Bigl\{a_1, f_1, w_1, n_1; a_2, f_2, w_2, n_2; \Omega_{b}h^2, \Omega_{c}h^2,  \nonumber \\
100\theta_{\rm MC},
  \tau, n_{s}, \ln[10^{10}A_{s}] \Bigr\}~.
\label{eq:two-step-parameters}
\end{eqnarray}
Subsequently, the set of free parameters representing parameter space for Three-Step class of models ($N=3$) is given by
\begin{eqnarray}
\mathcal{P}_3 \equiv\Bigl\{a_1, f_1, w_1, n_1; a_2, f_2, w_2, n_2; a_3, f_3, w_3, n_3; \nonumber \\
\Omega_{b}h^2, \Omega_{c}h^2, 100\theta_{\rm MC},
  \tau, n_{s}, \ln[10^{10}A_{s}] \Bigr\} .
\label{eq:three-step-parameters}
\end{eqnarray}

\section{Results}
\label{sec:results}

For testing our proposal we consider three main classes of models: \textbf{One-Step}, \textbf{Two-Step} and \textbf{Three-Step} models which in each class we assume there is a transition  in dark energy density. For simplicity we begin with One-Step models and continue to Two-Step and Three-Step class of models. We will test the effects of step-position $z_{{\rm c}, i}$, strength of fraction of total energy density $f_i$ and also the effects of having dS ($f_i>0$) or 
AdS ($f_i <0$) phases of evolution in the following subsections.

\subsection{One-Step models}
\label{sec:OneStep}

\subsubsection{Effects of step position}
In this subsection we consider One-Step models that are different in the position $z_{{\rm c},1}$ when the transition in dark energy takes place.   In Table.~ \ref{tab:onestep-priors-step-position} we see our considered priors for this class of models. When we let the step position ``$z_{{\rm c},1}$'' and the fraction of energy density ``$f_1$'' both to be free parameters we had some computational problems. Therefore,  we decided to test our hypothesis by assuming these two parameters to be fixed in each run but to have different values in each run. So, in this and the rest of our analysis we assume the step positions and the fraction of total energy density to be fixed parameters but to have different fixed values in each analysis. 
\begin{table*}[t]
\centering
\begin{tabular} {c||c c c c c}
Parameter & Priors  & Priors & Priors    \\
& Model 1  & Model 2 & Model 3   \\
\hline
\hline
{\boldmath${ 1+ z_{{\rm c},1}} $} & {~~ \boldmath $25$}  & {~~\boldmath $250$}  & {~~\boldmath $2500$}   \\
\hline
{$f_1$} & 0.20   & 0.20  & 0.20  \\
\hline
{$w_1$} & $[0.5, 1]$  & $[0.5, 1]$ & $[0.5, 1]$  \\
{$n_1$} & $[-1, 1.5]$  & $[-1, 1.5]$ &  $[-1, 1.5]$  \\
\hline
\end{tabular}
\caption{ Priors for \textbf{One-Step} models for different step positions.  Note that we take the  fraction of dark energy density $f_1$ for all cases to be the same 
while let $w_1$ and $n_1$ to vary. The redshift of the step position is denoted by $1+z_{\rm c, 1}$ and test the model in three different step positions: before recombination, after recombination and late-time era. }
\label{tab:onestep-priors-step-position}
\end{table*}

In Table.~\ref{tab:one-step-results-step-position} we summarize observational constrains of this class of models by considering $\Lambda$CDM as an anchor. We point out that in all of the analysis of models we use a combination of three types of data i.e. ``CMB+BAO+SN''.

Looking at Table.~\ref{tab:one-step-results-step-position} it is obvious that Model 3 shows the least tension with the SH0ES results. In this model we assume the transition in dark energy  occurs at $ z  \simeq 2500$ which is deeply before the surface of last scattering  and after the time of  matter-radiation equality. Our conclusion for {One-Step} models is that an early phase of dark energy with $f_1 >0$, as in EDE scenario, can reconcile the $H_0$ tension. 
On the other hand, {One-Step} models with transition happening after the surface of last scattering are not promising for this purpose.

\begin{table*}[t]
\centering
\begin{tabular} {c||c c c c c}
Parameter & Best-fit  & Best-fit & Best-fit & Best-fit \\
&$\Lambda$CDM & Model 1  & Model 2 & Model 3  \\
\hline
\hline
{\boldmath$n_1$} & $---$& $  > 0.945$  & $> 1.18$ & $> 1.46$  \\
\hline
{\boldmath$w_1 $} &$---$  & $> 0.880$ & $> 0.923$ & $0.808\pm 0.060$ \\
\hline

{\boldmath$\Omega_m $} &$0.3092\pm 0.0070$ & $0.3107\pm 0.0073$  & $0.3040\pm 0.0068$ & $0.2838\pm 0.0085$  \\
\hline
{\boldmath$H_0$} & $67.70\pm 0.52$&$ 66.43\pm 0.54$ & $67.80\pm 0.52$ & $71.83^{+0.59}_{-0.67}$  \\
\hline
{\boldmath$S_8 $} & $0.819\pm 0.014$& $0.793\pm 0.014$ & $0.803^{+0.013}_{-0.015}$ & $0.890\pm 0.018$ \\
\hline
{\boldmath$10^9 A_s $} & $2.091\pm 0.027$& $2.069^{+0.025}_{-0.028}$ & $2.086\pm 0.027$ & $2.052\pm 0.029$  \\
\hline
{\boldmath$n_s $} & $0.9668\pm 0.0042$ &$ 0.9795\pm 0.0044$ & $0.9681\pm 0.0042$ & $ 0.9932\pm 0.0044$  \\
\hline
{\boldmath$\tau $} & $0.0541\pm 0.0059$ & $0.0544\pm 0.0058$ & $0.0551\pm 0.0059$ & $0.0425\pm 0.0065$ \\
\hline
\hline

\end{tabular}
	\caption{\label{tab:one-step-results-step-position}  $\%$68 limits for parameters of \textbf{One-Step} models with different step positions (cf. table \ref{tab:onestep-priors-step-position}) in comparison with $\Lambda$CDM model based on CMB+BAO+SN data. }
\end{table*}

\begin{figure}[ht]
\centering
  \includegraphics[width=\linewidth]{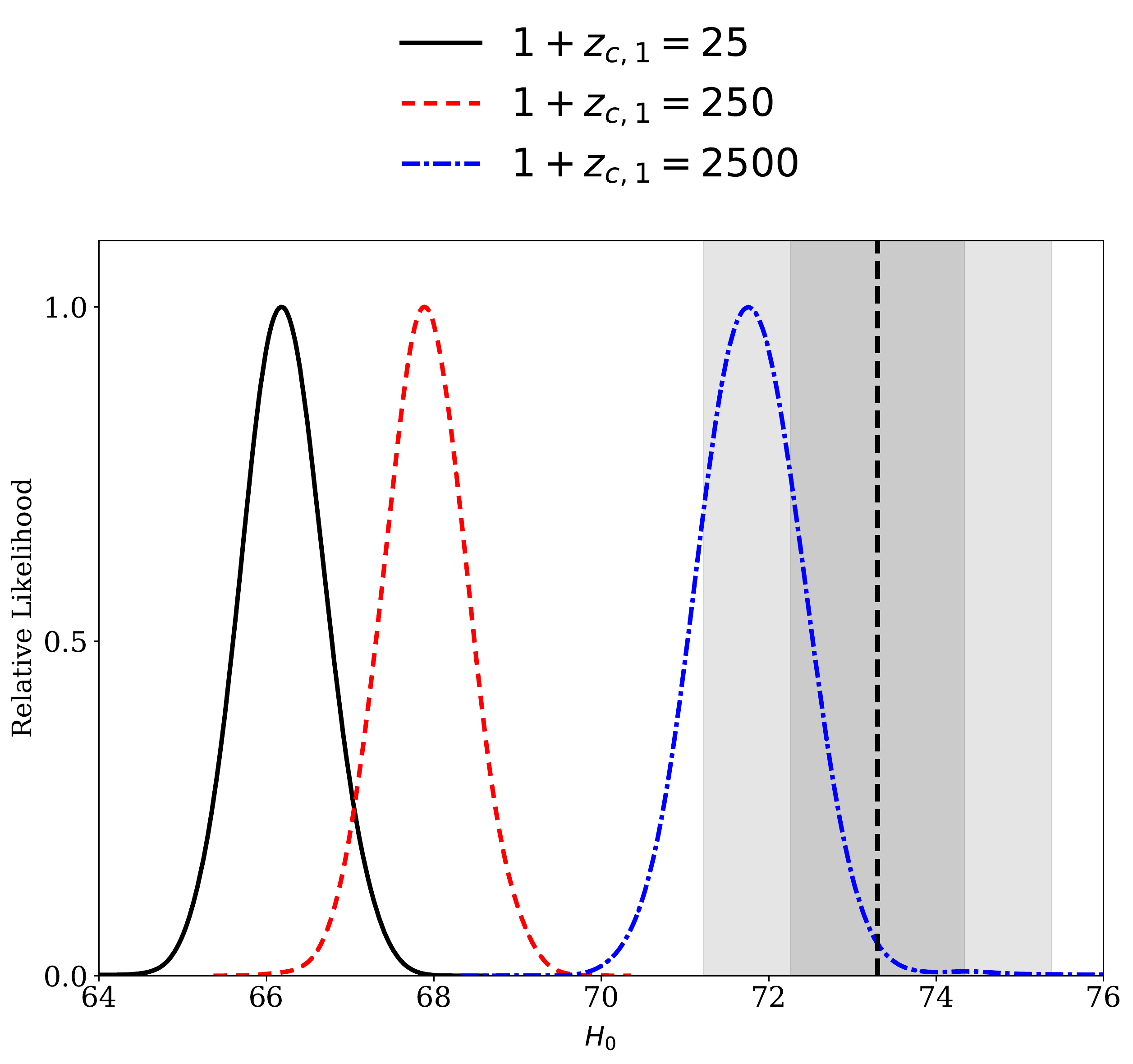}
  \caption{One-dimensional likelihoods for \textbf{One-Step} models for $H_0$  based on ``CMB+BAO+SN'' data. Note that the shaded area shows measurement of $H_0$ done by SH0ES team and its $1\sigma$ error ~\citep{Riess:2021jrx}.}
  \label{fig:one-step-H0-likelihoods-step-position}
\end{figure}
\begin{figure}[ht]
\centering
  \includegraphics[width=\linewidth]{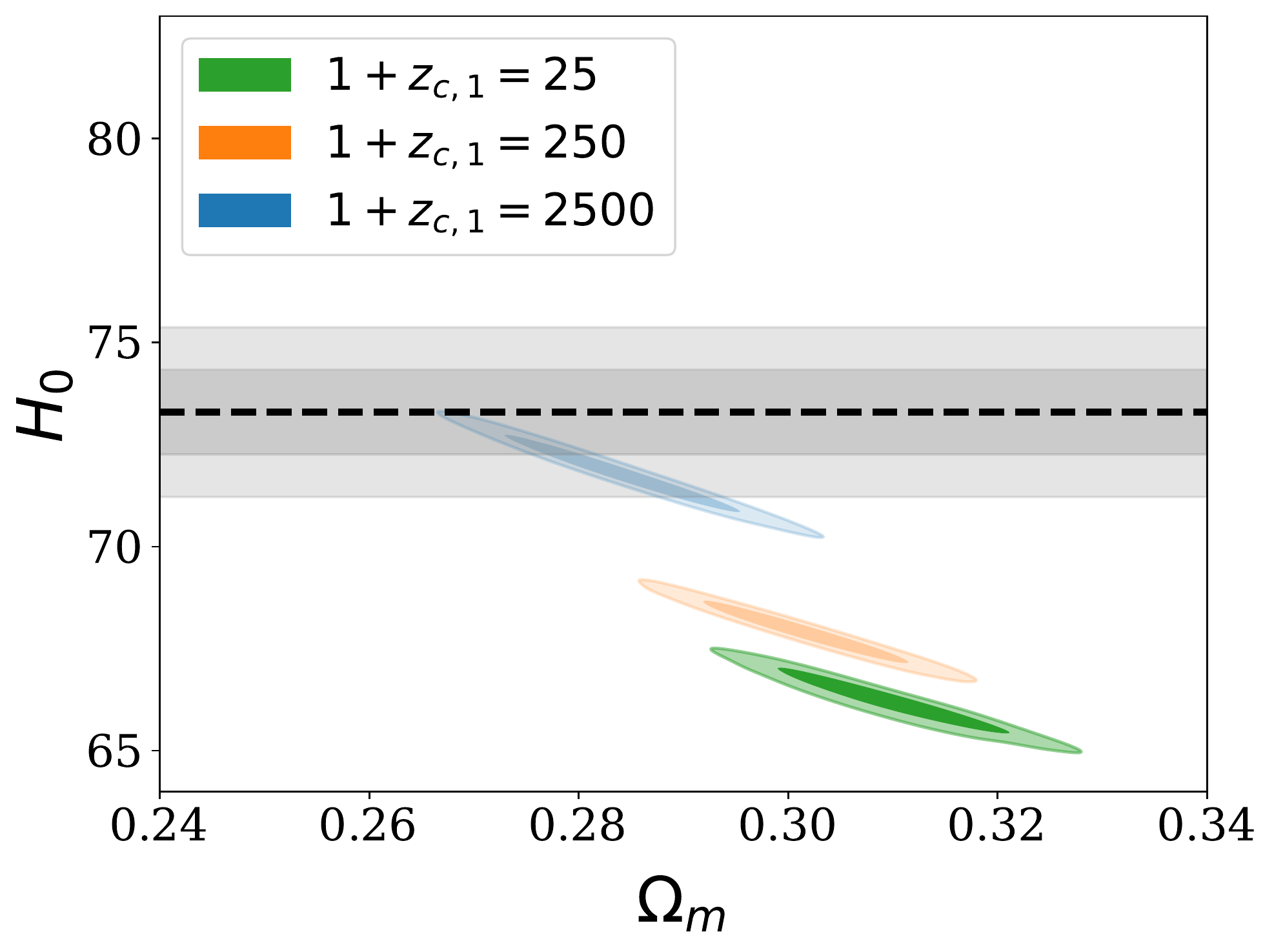}
  \caption{Contour plots for \textbf{One-Step} models for $H_0$ vs. $\Omega_m$ based on CMB+BAO+SN data. Note that the shaded area shows measurement of $H_0$ done by SH0ES team and its $1\sigma$ error \citep{Riess:2021jrx}.}
  \label{fig:one-step-H0-omegam-contours-step-position}
\end{figure}

The effects of considering early step in density evolution are shown in Fig.~\ref{fig:one-step-H0-likelihoods-step-position}. In this figure we see the likelihoods of different One-Step models with different step-positions. Also, contour plots for parameters $H_0$ and $\Omega_m$ can be seen in Fig.~\ref{fig:one-step-H0-omegam-contours-step-position}.

\subsubsection{Effects of the strength of the fraction of dark energy density}
In this section we look at the effects of the fraction of dark energy density in One-Step class of model. We summarize our consideration for priors in this analysis in Table.~\ref{tab:one-step-priors-f-strength}. According to our past analysis we choose a fixed step position at $a_{{\rm c},1}=0.0004~(1+z_{{\rm c},1}=2500)$ and also fixed the fraction of dark energy density $f_1$ but with different values in each analysis. Both of $w_1$ and $n_1$ are free parameters in all of our analysis.

\begin{figure}[ht]
\centering
  \includegraphics[width=\linewidth]{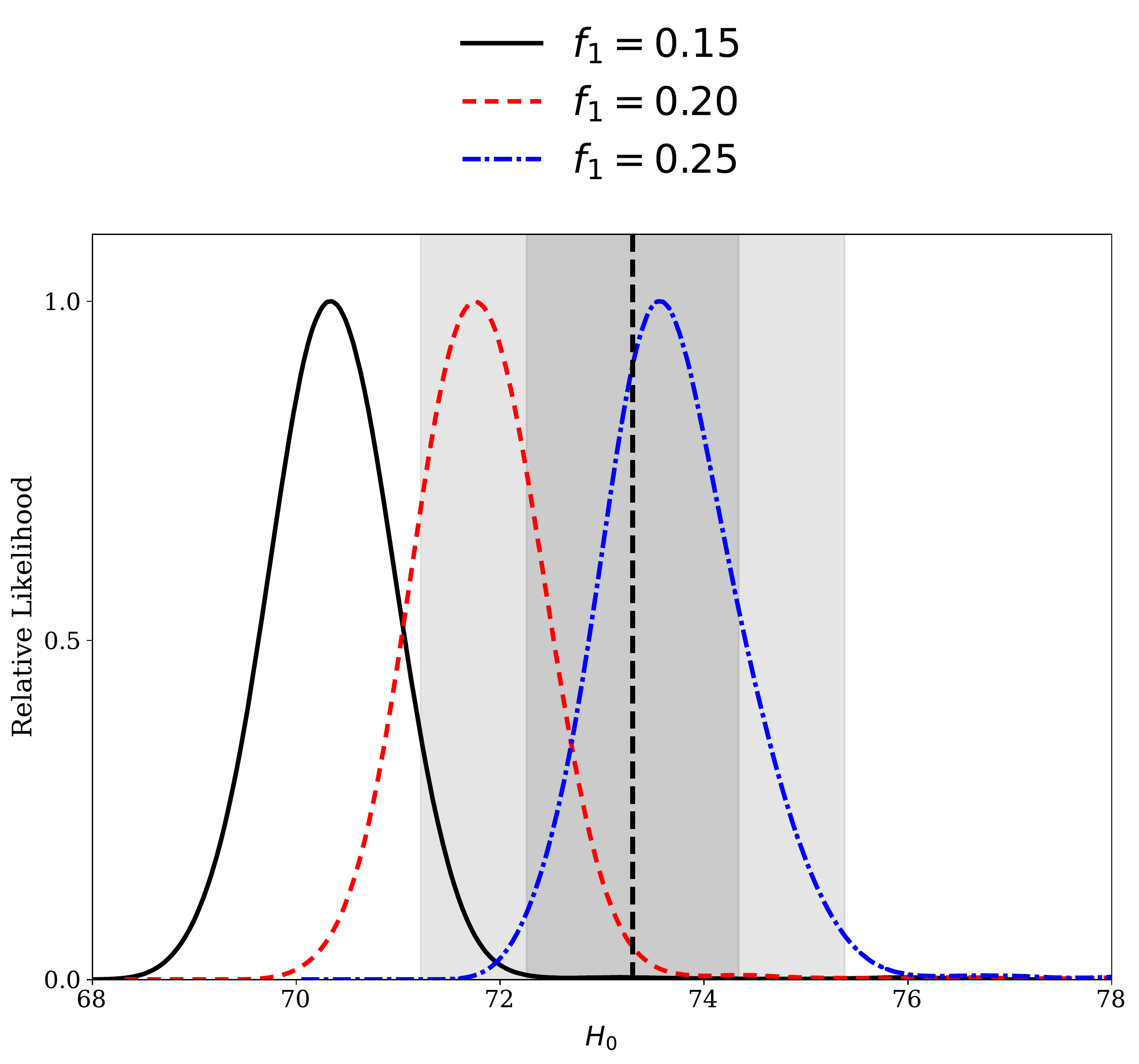}
  \caption{\footnotesize One-dimensional likelihoods for \textbf{One-Step} models with different values of $f_1$ for $H_0$  based on CMB+BAO+SN data.
  Note that the shaded area shows measurement of $H_0$ done by SH0ES team and its $1\sigma$ error~\citep{Riess:2021jrx}.}
  \label{fig:one-step-H0-likelihoods-f-strength}
\end{figure}

\begin{figure}[ht]
\centering
  \includegraphics[width=\linewidth]{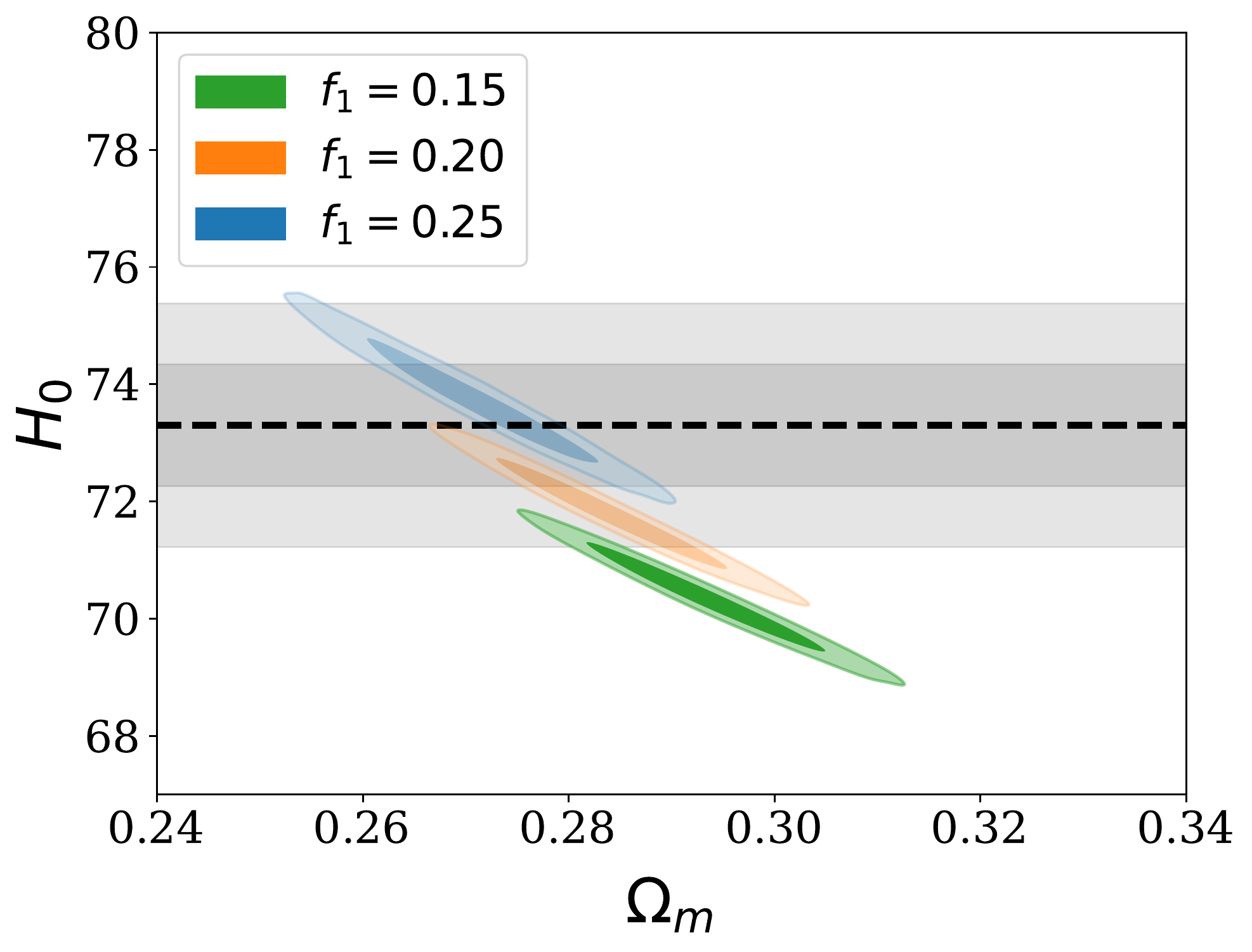}
  \caption{\footnotesize Contour plots for \textbf{One-Step} models with different values of $f_1$ for $H_0$ vs. $\Omega_m$ based on CMB+BAO+SN data. Note that the shaded area shows measurement of $H_0$ done by SH0ES team and its $1 \sigma$ error \citep{Riess:2021jrx}.}
  \label{fig:one-step-H0-omegam-contours-f-strength}
\end{figure}

\begin{table*}[ht]
\centering
\begin{tabular} {c||c c c c c}
Parameter & Priors  & Priors & Priors    \\
& Model 4  & Model 3 & Model 5   \\
\hline
\hline
{${ 1+z_{{\rm c},1}}$} & {~~ 2500}  & {~~ 2500} & {~~ 2500}   \\
\hline
{\boldmath ${f_1}$} & {\boldmath $0.15$}  & {\boldmath $0.20$} & {\boldmath $0.25$}  \\
\hline
{$w_1$} & $[0.5, 1]$  & $[0.5, 1]$ & $[0.5, 1]$  \\
{$n_1$} & $[-1, 1.5]$  & $[-1, 1.5]$ &  $[-1, 1.5]$  \\
\hline
\end{tabular}
\caption{Priors for \textbf{One-Step} models for different $f_1$ strengths.  We set the step position to be fixed before recombination era $1+z_{\rm c, 1}=2500$ and let the  fraction of energy density $f_1$ varies for different cases.}
\label{tab:one-step-priors-f-strength}
\end{table*}

The conclusion is that the higher is the value of $f_1$, the larger is the prediction for $H_0$. 
As we see in Table.~\ref{tab:one-step-results-f-strength}  Model 5 which has higher value of fraction of energy density ($f_1=0.25$) shows more consistency with the value of $H_0$ from SH0ES measurements. As we reported in Table.~\ref{tab:H0-tension-criteria} the Gaussian tension in $H_0$ for model 5 is just $0.47 \sigma$.

\begin{table*}[ht]
\centering
\begin{tabular} {c||c c c c c}
Parameter & Best-fit  & Best-fit & Best-fit \\
& Model 4  & Model 3 & Model 5 \\
\hline
\hline
{\boldmath$n_1$} & $  > 1.42$  &$> 1.46$  & $> 1.47$  \\
\hline
{\boldmath$w_1 $} &  $0.787^{+0.064}_{-0.092}$ & $0.808\pm 0.060$  & $0.820^{+0.073}_{-0.032}$\\
\hline
{\boldmath$\Omega_m $} & $0.2933\pm 0.0080$  & $0.2838\pm 0.0085$ &  $0.2703^{+0.0089}_{-0.0061}$  \\
\hline
{\boldmath$H_0$} & $ 70.37\pm 0.68$ & $71.83^{+0.59}_{-0.67}$ & $73.84^{+0.49}_{-0.92}$  \\
\hline
{\boldmath$S_8 $} & $0.873\pm 0.017$ & $0.890\pm 0.018$ & $0.903^{+0.018}_{-0.014}$  \\
\hline
{\boldmath$10^9 A_s $} & $2.058\pm 0.027$ & $2.052\pm 0.029$ & $2.053\pm 0.030$   \\
\hline
{\boldmath$n_s $} & $  0.9815\pm 0.0042$ & $ 0.9932\pm 0.0044$ & $1.0098\pm 0.0055$  \\
\hline
{\boldmath$\tau $} & $0.0449\pm 0.0060$ & $0.0425\pm 0.0065$ & $0.0414^{+0.0063}_{-0.0077}$ \\

\hline
\hline

\end{tabular}
	\caption{\label{tab:one-step-results-f-strength} Observational constraints and $\%$68 limits for parameters of \textbf{One-Step} models with different values for $f_1$ (cf. table \ref{tab:one-step-priors-f-strength}) based on CMB+BAO+SN data.
The higher is the value of $f_1$, the larger is the prediction for $H_0$. }
\end{table*}

In Fig.~\ref{fig:one-step-H0-likelihoods-f-strength} we present the likelihood probabilities for One-Step class of models with different values for $f_1$. Also, the two-dimensional contour plots for parameters $H_0$ vs. $\Omega_m$ are shown in Fig.~\ref{fig:one-step-H0-omegam-contours-f-strength}. In both figures the shaded area are $1 \sigma$ and $2 \sigma$ allowed error regions based on SH0ES measurements for $H_0$ \citep{Riess:2021jrx}.

\vspace{1cm}

\subsection{Two-Step models} 
\label{sec:Two-Step models}

Here, we extend our analysis to some Two-Step class of models. In this type of models the first transition in dark energy can occur in more earlier time while the second transition can take place before the surface of last scattering or after that. Also, we have freedom in choosing the second phase of dark energy 
to be dS type or AdS type.

\vspace{0.5cm}
\subsubsection{Effect of position of second step}

In this approach we assume the first step is fixed at the time $1+z_{\rm c, 1}=2500$ and test different positions for the second step of evolution. In Table.~\ref{tab:two-step-priors-second-step-position} we present our assumptions for priors of parameters. Since we consider the first step to occur before the surface of last scattering,  we assume the second step to occur after the surface of last scattering. In addition,  in all of the models in this subsection we assume a dS phase for second step (positive values for $f_2$).
\begin{table*}
\centering
\begin{tabular} {c||c c c c c}
Parameter & Priors  & Priors  & Priors  \\
& Model 6 & Model 7 & Model 8 \\
\hline
\hline
{$1+z_{{\rm c},1}$} & 2500  & 2500 & 2500  \\
\hline
{$f_1$} & 0.20  & 0.20  & 0.20    \\
\hline
{$w_1$} & $[0.5, 1]$  & $[0.5, 1]$ & $[0.5, 1]$  \\
\hline
{$n_1$} & $[-1, 1.5]$ & $[-1, 1.5]$ & $[-1, 1.5]$  \\
\hline
{\boldmath$1+z_{{\rm c},2}$} & {\boldmath $833$}  & {\boldmath $83.3$}  & {\boldmath $8.3$}  \\
\hline
{$f_2$} & 0.10  & 0.10 & 0.10  \\
\hline
{$w_2$} & $[0.5, 1]$  & $[0.5, 1]$ & $[0.5, 1]$  \\
\hline
{$n_2$} & $ [-1, 1.5]$ & $[-1, 1.5]$ & $[-1, 1.5]$  \\
\hline

\end{tabular}
\caption{Priors for \textbf{Two-Step} models with different second step positions $1+z_{{\rm c},2}$ while all other inputs are the same. We choose the first step to occur  before recombination era $1+z_{{\rm c},1}=2500$ and let the second step  to occur after recombination but in three different era.}
\label{tab:two-step-priors-second-step-position}
\end{table*}

Looking at Table.~\ref{tab:two-step-results-second-step-position} we see that 
the result for $H_0$ does not strongly depend 
on the position of the second step (as long as the sign of $f_2$ is fixed). For example in the the  Models 6, 7 and 8, the values of $z_{2, \rm c}$ change by two orders of magnitude while $H_0$ does not change drastically. Having said this, we notice that the value of $H_0$ in Model 6, where the second step occurs more closer to CMB era,  shows more consistency with the SH0ES measurements. On the other hand, by  putting the second step closer to late-time (Model 8) $H_0$ shows more tension.


\begin{table*}
\centering
\begin{tabular} {c||c c c c c}
Parameter & Best-fit & Best-fit& Best-fit \\
& Model 6 & Model 7 & Model 8 \\
\hline
\hline
{\boldmath$n_1$} & $ > 1.47$  & $> 1.42$ & $ > 1.46$ \\
\hline
{\boldmath$w_1 $} & $ > 0.984$ & $  0.763^{+0.036}_{-0.057}$ & $ 0.703^{+0.033}_{-0.043}$  \\
\hline
{\boldmath$n_2$} & $ > 0.363$  & $---$ & $---$ \\
\hline
{\boldmath$w_2$} & $ > 0.730$ & $< 0.641$ & $ < 0.576$  \\
\hline
{\boldmath$\Omega_m $} & $0.2801^{+0.0081}_{-0.0055}$  & $0.2811^{+0.0085}_{-0.0068}$ &$0.2832\pm 0.0076$   \\
\hline
{\boldmath$H_0$} & $71.92^{+0.39}_{-0.74}$ & $71.55^{+0.52}_{-0.77}$ & $ 70.11^{+0.59}_{-0.76}$  \\
\hline
{\boldmath$S_8 $} & $0.876^{+0.017}_{-0.013}$ & $ 0.875\pm 0.021$ & $0.853\pm 0.017$  \\
\hline
{\boldmath$10^9 A_s $} & $ 2.019\pm 0.031$ & $2.039^{+0.025}_{-0.036}$ & $ 2.017^{+0.026}_{-0.030}$ \\
\hline

{\boldmath$n_s $} & $ 0.9843\pm 0.0048$ & $0.9989^{+0.0050}_{-0.0044}$  & $1.0085^{+0.0052}_{-0.0044}$ \\
\hline
{\boldmath$\tau $} & $ 0.0426\pm 0.0072$ & $0.0434^{+0.0058}_{-0.0078}$ & $ 0.0448^{+0.0056}_{-0.0063}$  \\
\hline
\hline

\end{tabular}
	\caption{\label{tab:two-step-results-second-step-position} $\%$68 limits for parameters of \textbf{Two-Step} models based on CMB+BAO+SN data. Note that all the input parameters are the same and only the position of the second step $z_{\rm c, 2}$ is different (cf. table \ref{tab:two-step-priors-second-step-position}).
While $z_{\rm c, 2}$ changes by two orders of magnitude in these examples, but $H_0$ does not change significantly. 	
}
\end{table*}

\begin{figure}[ht]
\centering
  \includegraphics[width=\linewidth]{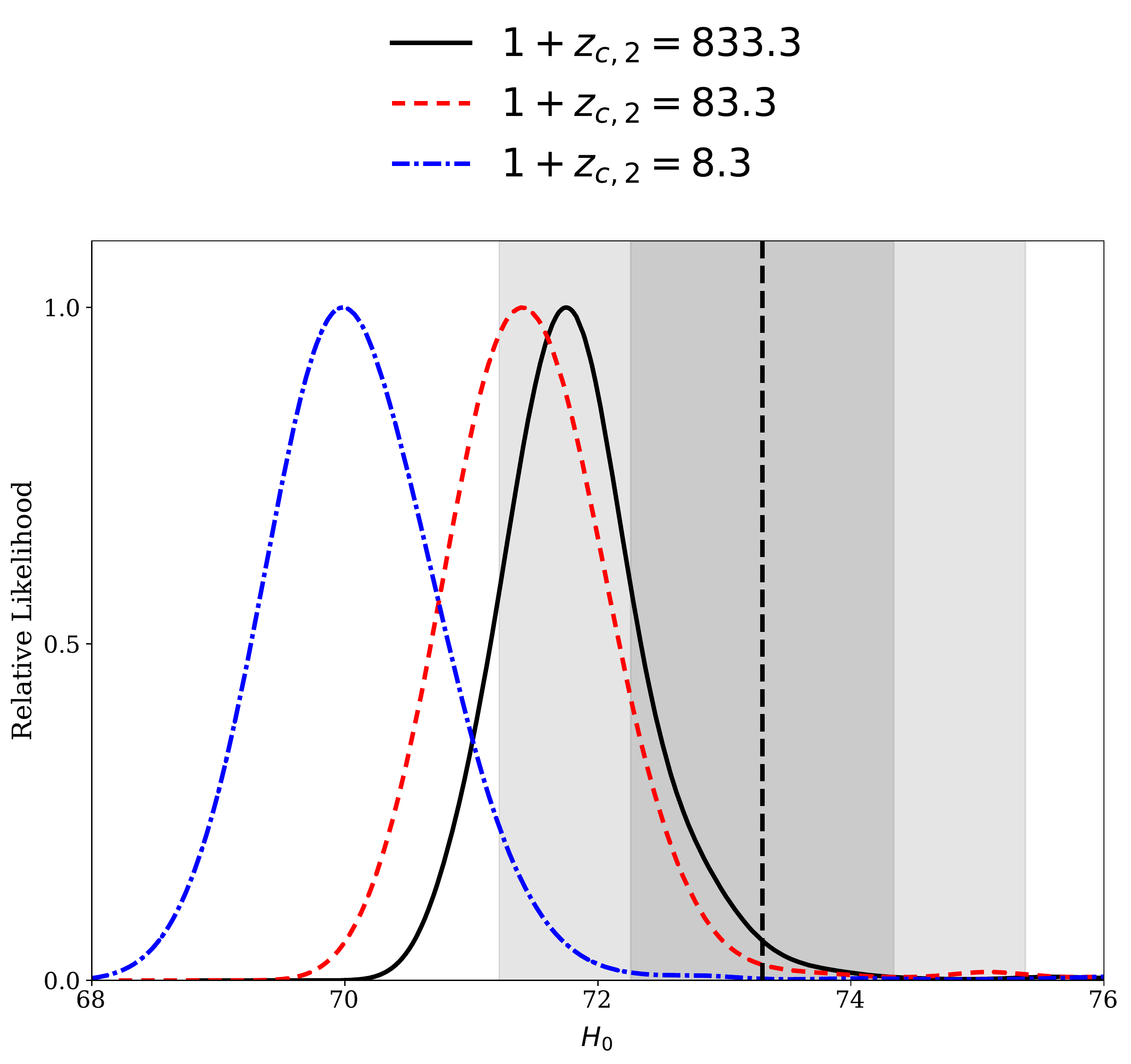}
  \caption{\footnotesize One-dimensional likelihoods for \textbf{Two-Step} models for $H_0$  based on CMB+BAO+SN data comparing the effect of second step position $z_{\rm c, 2}$ with fixed $f_1, f_2 >0$. }
  \label{fig:two-step-H0-likelihoods-second-step-effect}
\end{figure}

\begin{figure}[ht]
\centering
  \includegraphics[width=\linewidth]{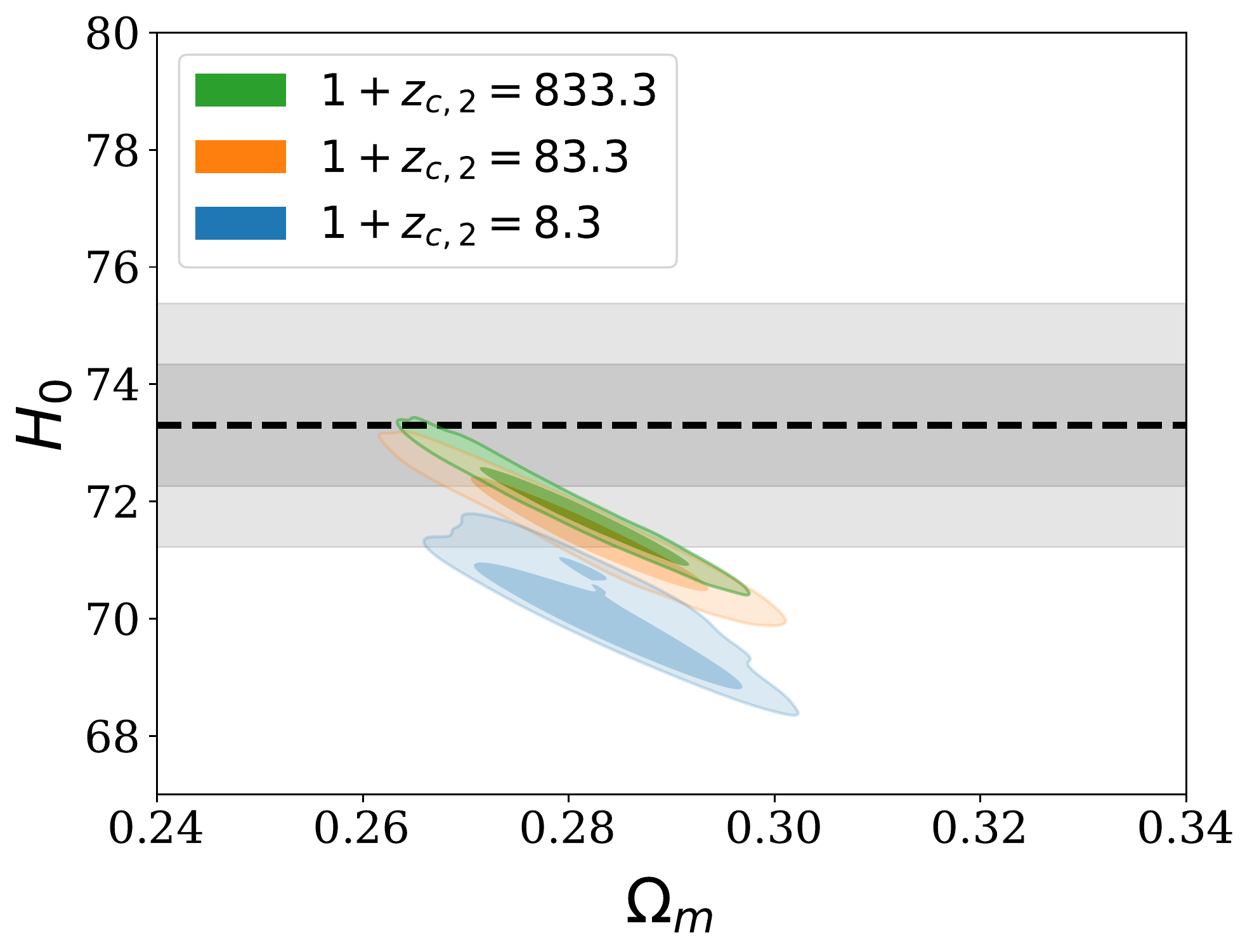}
  \caption{\footnotesize Contour plots for \textbf{Two-Step} models for $H_0$ vs. $\Omega_m$ based on CMB+BAO+SN data comparing the effect of the second step position $z_{\rm c, 2}$ with fixed $f_1, f_2 >0$.  Note that the shaded area shows measurement of $H_0$ done by SH0ES team and its $1 \sigma$ error \citep{Riess:2021jrx}.}
  \label{fig:two-step-H0-omegam-contours-second-step-effect}
\end{figure}

In Figs.~\ref{fig:two-step-H0-likelihoods-second-step-effect} and~\ref{fig:two-step-H0-omegam-contours-second-step-effect} we present the likelihoods and contour plots for $H_0$ and $\Omega_m$ in Two-Step models with different position of second step.

\subsubsection{Considering dS or AdS for second step}

As mentioned above in Two-Step and Three-Step models we have the freedom in choosing the next phases to be dS-like or AdS-like. Hence, we want to test the effect of having a dS or an AdS in the second phase of evolution. In Table.~\ref{tab:two-step-priors-dS-AdS} we have priors for two models: Model 7 and Model 9 that have similar priors but with opposite signs for $f_2$.


\begin{table*}[ht]
\centering
\begin{tabular} {c||c c c c c}
Parameter & Priors  & Priors    \\
& Model 7 & Model 9 \\
\hline
\hline
{$1+z_{{\rm c},1}$} & 2500 & 2500   \\
\hline
{$f_1$} &  0.20 & 0.20    \\
\hline
{$w_1$} &  $[0.5, 1]$ & $[0.5, 1]$  \\
\hline
{$n_1$} &  $[-1, 1.5]$ & $[-1, 1.5]$  \\
\hline
{$1+z_{{\rm c},2}$} &  83.3  & 83.3  \\
\hline
{\boldmath$f_2$} &  {\boldmath $+0.10$} & {\boldmath $-0.10$}   \\
\hline
{$w_2$} &  $[0.5, 1]$ & $[0.5, 1]$  \\
\hline
{$n_2$} & $[-1, 1.5]$ & $[-1, 1.5]$  \\
\hline
\end{tabular}
\caption{Priors for \textbf{Two-Step} models with different signs for $f_2$.  As in standard EDE scenario, we choose the first step to be a dS type $f_1 > 0$ while let the second step to be either 
dS type $(f_2 >0)$ or AdS type $(f_2 <0)$. Other parameters are the same.}
\label{tab:two-step-priors-dS-AdS}
\end{table*}

\begin{figure}[h]
\centering
  \includegraphics[width=\linewidth]{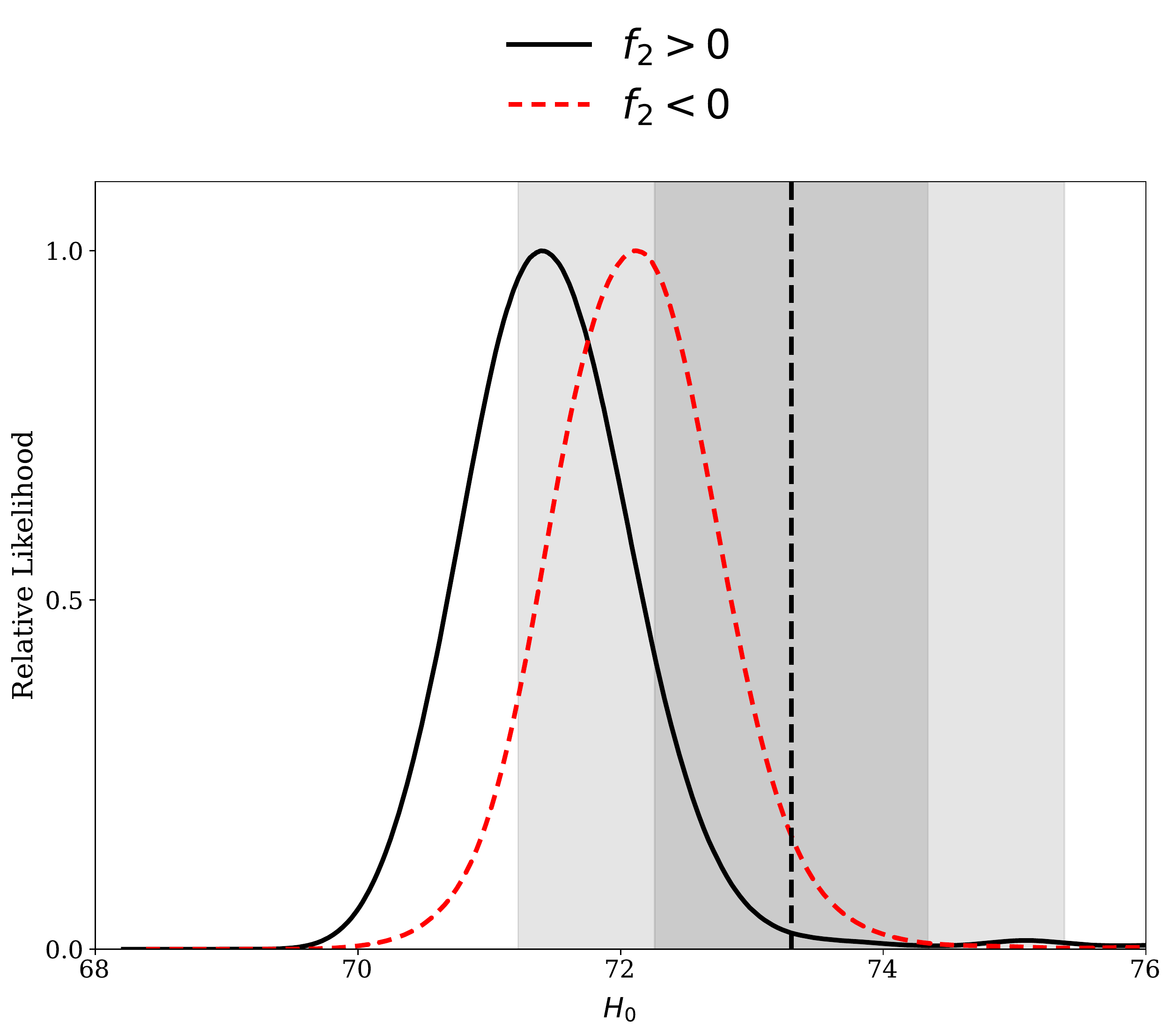}
  \caption{One-dimensional likelihoods for \textbf{Two-Step} models for $H_0$  based on CMB+BAO+SN data comparing the effect of the sign of $f_2$, dS ($f_2>0$) or AdS ($f_2 <0$) in Models 7 and 9.}
  \label{fig:two-step-H0-likelihoods-ds-ads}
\end{figure}

\begin{figure}[h]
\centering
  \includegraphics[width=\linewidth]{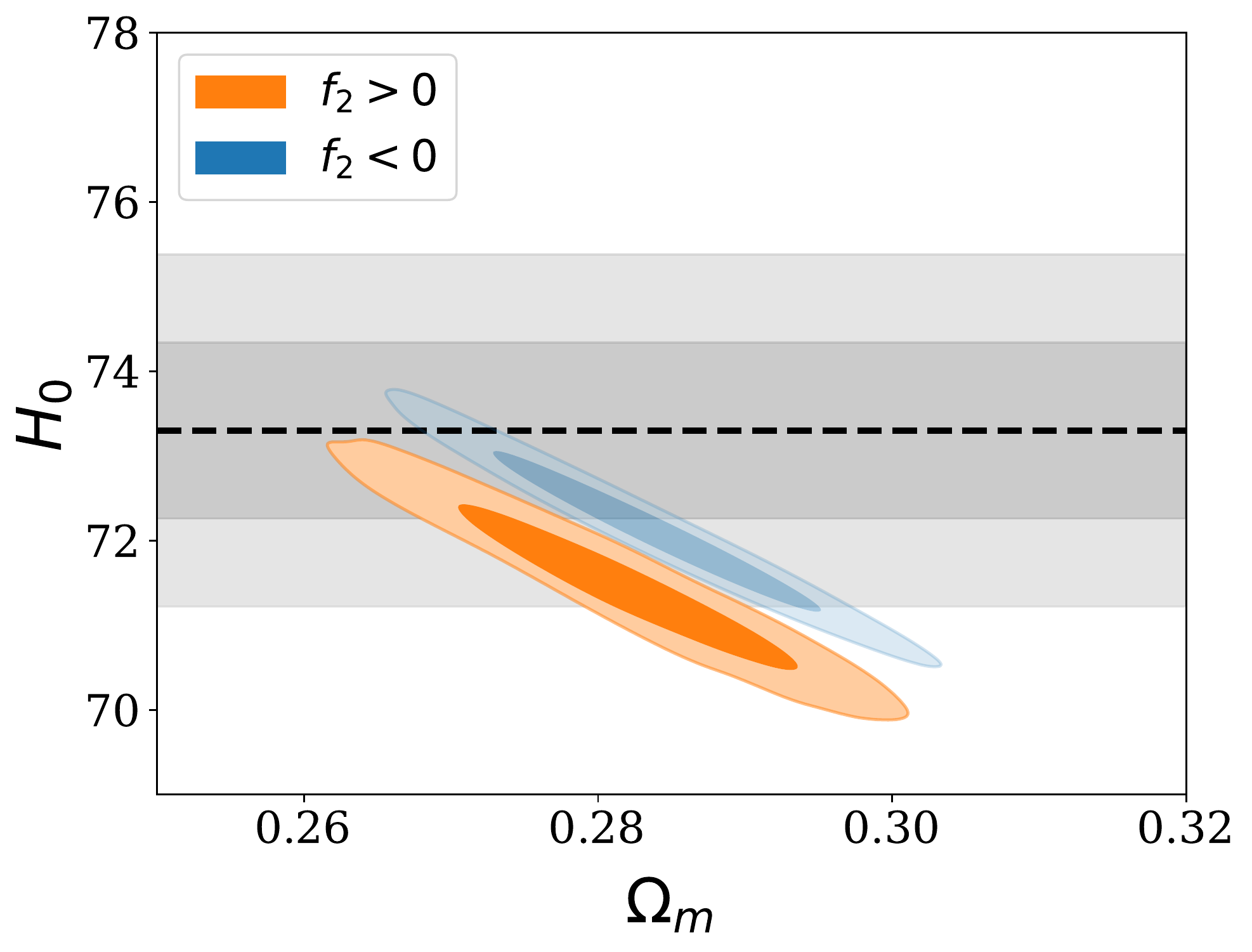}
  \caption{\footnotesize Contour plots for \textbf{Two-Step} models for $H_0$ vs. $\Omega_m$ based on CMB+BAO+SN data comparing the effect of the sign of $f_2$,   dS ($f_2>0$) or AdS ($f_2 <0$) in Models 7 and 9.  Note that the shaded area shows the measurement of $H_0$ done by SH0ES team and its $1 \sigma$ error \citep{Riess:2021jrx}.}
  \label{fig:two-step-H0-omegam-contours-ds-ads}
\end{figure}

In Table.~\ref{tab:two-step-results-dS-AdS} we summarize observational constrains for parameters of Model 7 and Model 9 while in Figs.~\ref{fig:two-step-H0-likelihoods-ds-ads} and ~\ref{fig:two-step-H0-omegam-contours-ds-ads} we see likelihoods and contour plots for $H_0$ and $\Omega_m$ in these two models. 
Clearly, having the second phase to be AdS-type, yields to a higher value of $H_0$. Furthermore, in this numerical example the second phase to be AdS-type
shows more consistency with the SH0ES measurements.

\begin{figure}[h]
\centering
  \includegraphics[width=\linewidth]{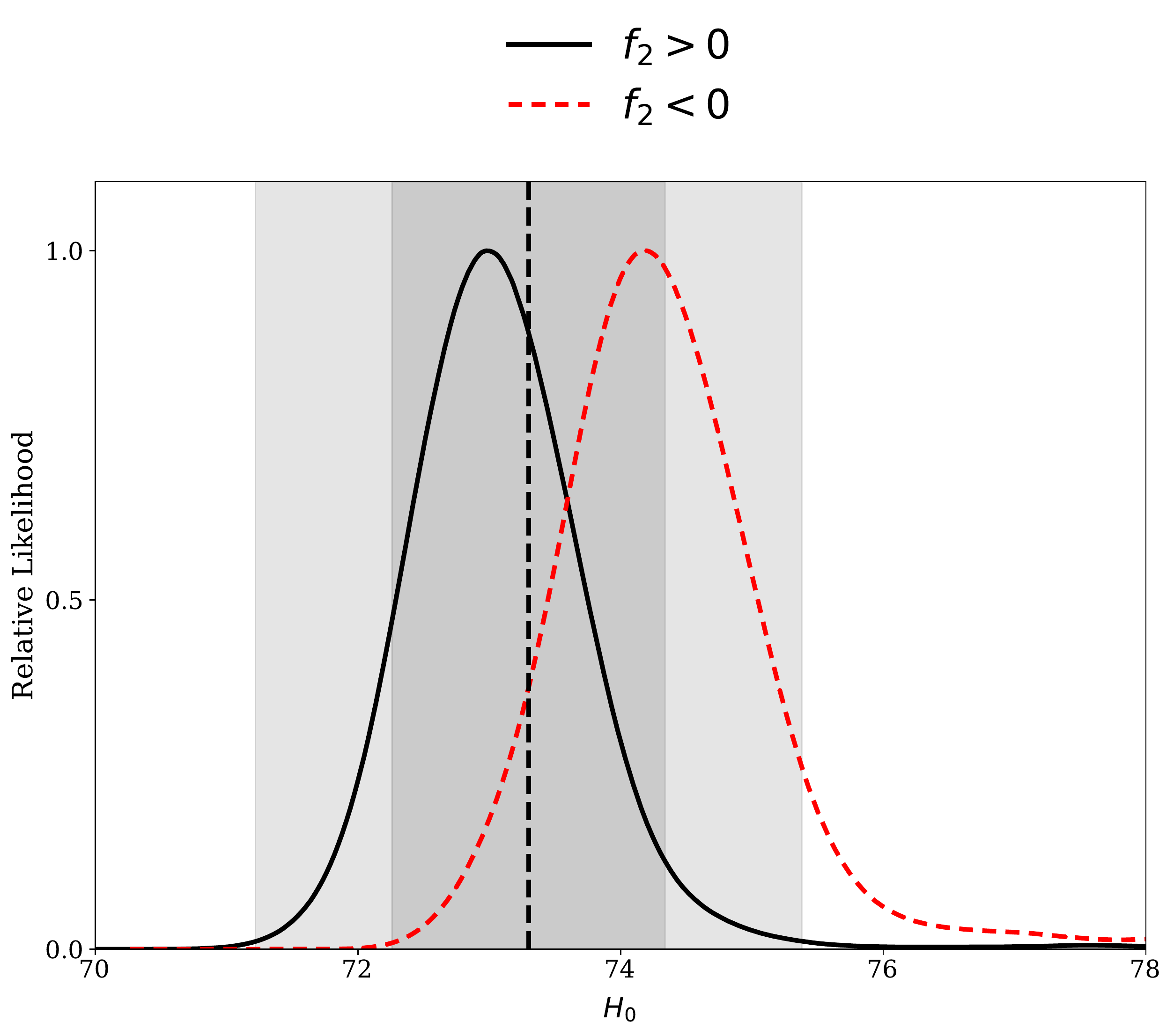}
  \caption{One-dimensional likelihoods for \textbf{Two-Step} models for $H_0$  based on CMB+BAO+SN data, comparing the effects of 
  the sign of $f_2$ in Models 10 and 11. This plot is parallel to Fig. \ref{fig:two-step-H0-likelihoods-ds-ads} performed for Models 7 and 9 but now the value of $f_1$ is increased to $f_1=0.25$ and the values of $f_2$ in dS and AdS cases are changed to $f_2=+0.15$ and $f_2=-0.15$ respectively. }
  \label{fig:two-step-H0-likelihoods-ds-ads-higher}
\end{figure}

\begin{figure}[h]
\centering
  \includegraphics[width=\linewidth]{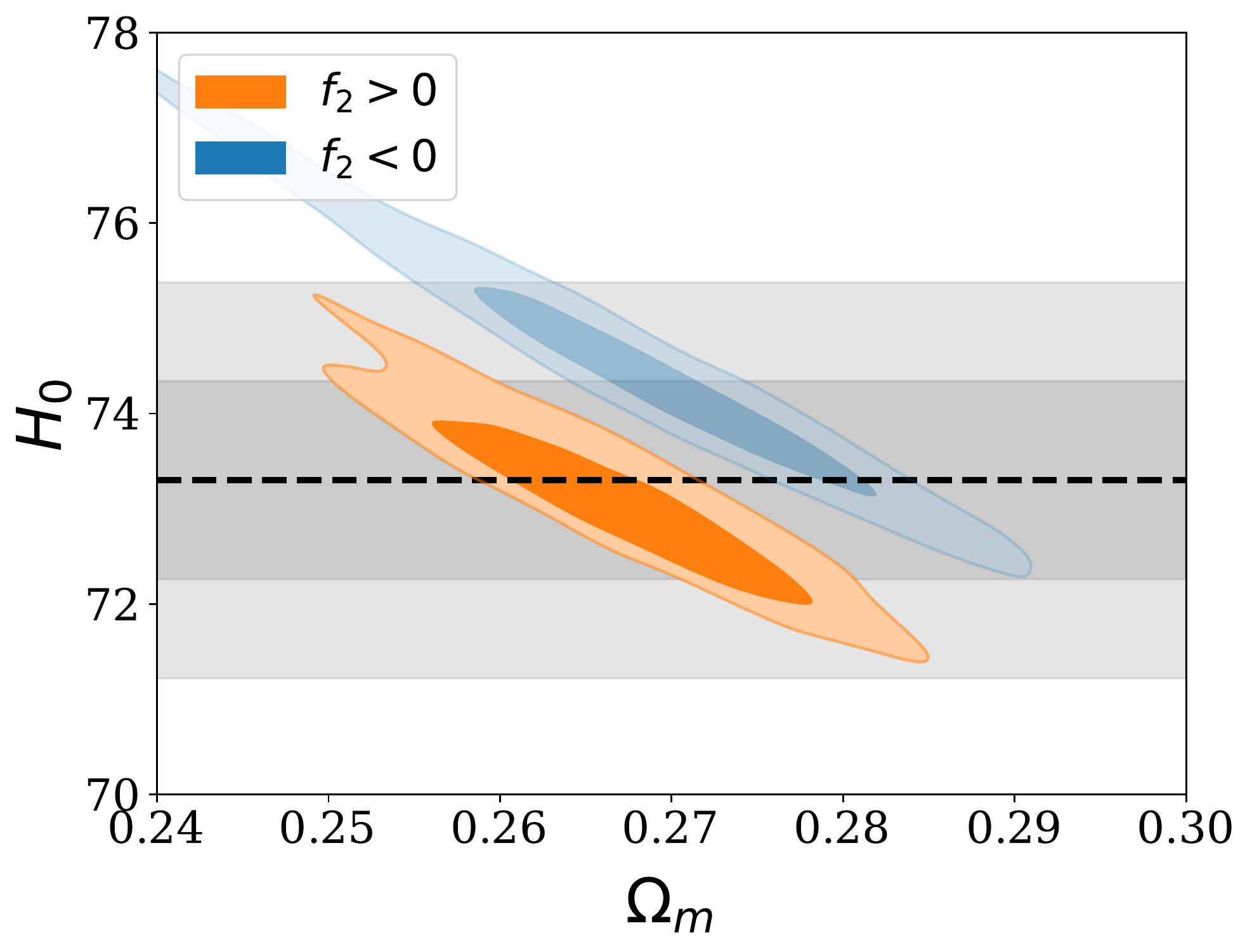}
  \caption{Contour plots for \textbf{Two-Step} models for $H_0$ vs. $\Omega_m$ based on CMB+BAO+SN data comparing the effects of 
  sign of $f_2$. The rest of the description is as in Fig. \ref{fig:two-step-H0-likelihoods-ds-ads-higher}. }
  \label{fig:two-step-H0-omegam-contours-ds-ads-higher}
\end{figure}

To confirm these conclusions, we have repeated this comparison for the Models 10 and 11 as well which have opposite signs of $f_2$,  with the priors given in 
Table.~\ref{tab:twostep-priors-3} while the results are summarized in Table.~\ref{tab:twostep-results-3}.  In Figs.  \ref{fig:two-step-H0-likelihoods-ds-ads-higher}
and \ref{fig:two-step-H0-omegam-contours-ds-ads-higher} we see likelihoods and contour plots for $H_0$ and $\Omega_m$ in these two models. As expected, a second phase in AdS-type yields to a higher value of $H_0$.


\begin{table*}[ht]
\centering
\begin{tabular} {c||c c c c c}
Parameter & Best-fit & Best-fit  \\
&  Model 7 & Model 9 \\
\hline
\hline
{\boldmath$n_1$} &  $ > 1.42$ & $ > 1.46$ \\
\hline
{\boldmath$w_1 $} &  $0.763^{+0.036}_{-0.057}$ & $ 0.827\pm 0.061$  \\
\hline
{\boldmath$n_2$} & $ ---$ & $> 0.783$ \\
\hline
{\boldmath$w_2$} &  $< 0.641$ & $ > 0.794$  \\
\hline
{\boldmath$\Omega_m $} &  $0.2811^{+0.0085}_{-0.0068}$ &$0.2832^{+0.0078}_{-0.0070}$   \\
\hline

{\boldmath$H_0$} &  $71.55^{+0.52}_{-0.77}$ & $ 72.19^{+0.54}_{-0.72}$  \\
\hline
{\boldmath$S_8 $} &  $ 0.875\pm 0.021$ & $0.895^{+0.017}_{-0.015}$  \\
\hline
{\boldmath$10^9 A_s $} &  $2.039^{+0.025}_{-0.036}$ & $ 2.060\pm 0.029$ \\
\hline

{\boldmath$n_s $} &  $0.9989^{+0.0050}_{-0.0044}$  & $0.9914\pm 0.0046$ \\
\hline
{\boldmath$\tau $} &  $0.0434^{+0.0058}_{-0.0078}$ & $ 0.0427\pm 0.0065$  \\
\hline
\hline

\end{tabular}
	\caption{\label{tab:two-step-results-dS-AdS}  Best-fit values and $\%$68 confidence intervals for parameters of \textbf{Two-Step} models based on CMB+BAO+SN data. Everything is similar in two cases but sign of $f_2$ is opposite (cf. table \ref{tab:two-step-priors-dS-AdS}).}
\end{table*}


\begin{table*}[ht]
\centering
\begin{tabular} {c||c c c c c}
Parameter & Priors  & Priors    \\
& Model 10 & Model 11 \\
\hline
\hline
{$1+z_{{\rm c},1}$} & 2500  & 2500   \\
\hline
{$f_1$} &  0.25 & 0.25   \\
\hline
{$w_1$} &  $[0.5, 1]$ & $[0.5, 1]$  \\
\hline
{$n_1$} &  $[-1, 1.5]$ & $[-1, 1.5]$  \\
\hline
{$1+z_{{\rm c},2}$} &  83.3 & 83.3  \\
\hline
{\boldmath$f_2$} &  {\boldmath $0.15$} & {\boldmath $-0.15$}  \\
\hline
{$w_2$} &  $[0.5, 1]$ & $[0.5, 1]$  \\
\hline
{$n_2$} & $[-1, 1.5]$ & $[-1, 1.5]$  \\
\hline
\end{tabular}
\caption{Priors for \textbf{Two-Step} models with opposite sign of $f_2$. This  is similar to Table~\ref{tab:two-step-priors-dS-AdS} for Models 7 and 9 but here  we have increased  both $f_1$ and $|f_2|$ compared to  Table~\ref{tab:two-step-priors-dS-AdS}. }  
\label{tab:twostep-priors-3}
\end{table*}


\begin{table*}[ht]
\centering
\begin{tabular} {c||c c c c c}
Parameter & Best-fit & Best-fit  \\
&  Model 10 & Model 11 \\
\hline
\hline
{\boldmath$n_1$} &  $ > 1.46$ & $ > 1.41$ \\
\hline
{\boldmath$w_1 $} &  $0.762\pm 0.036$ & $ 0.824^{+0.087}_{-0.044}$  \\
\hline
{\boldmath$n_2$} & $ < 0.957$ & $> 1.07$ \\
\hline
{\boldmath$w_2$} &  $< 0.534$ & $ > 0.786$  \\
\hline
{\boldmath$\Omega_m $} &  $ 0.2662^{+0.0078}_{-0.0061}$ &$0.267^{+0.011}_{-0.0052}$   \\
\hline

{\boldmath$H_0$} &  $73.18^{+0.48}_{-0.81}$ & $ 74.70^{+0.34}_{-1.2}$  \\
\hline
{\boldmath$S_8 $} &  $ 0.875^{+0.017}_{-0.015}$ & $0.904^{+0.025}_{-0.013}$  \\
\hline
{\boldmath$10^9 A_s $} &  $2.021^{+0.026}_{-0.032}$ & $  2.060\pm 0.031$ \\
\hline

{\boldmath$n_s $} &  $1.0223^{+0.0054}_{-0.0047}$  & $1.0077\pm 0.0069$ \\
\hline
{\boldmath$\tau $} &  $0.0414^{+0.0059}_{-0.0067}$ & $ 0.0420^{+0.0061}_{-0.0087}$  \\
\hline
\hline

\end{tabular}
	\caption{\label{tab:twostep-results-3}  Summary of observational constraints and $\%$68 limits for parameters of \textbf{Two-Step} models based on CMB+BAO+SN data comparing effects of sign of $f_2$, dS or AdS (cf. table \ref{tab:twostep-priors-3}, 
	similar to table.~\ref{tab:two-step-results-dS-AdS} for Models 7 and 9). }
\end{table*}


\subsection{Three-Step models}
\label{sec:ThreeStep}

As a wider extension of our study we extend the analysis to the models with three steps of transitions in dark energy. As before, we assume the first phase is dS-type happening at early time at $z_{{\rm c},1} = 2500$ while the second and thirst steps can happen near or long after the surface of last scattering with either dS or AdS type.  As described in Table.~\ref{tab:three-step-priors-1},  we have tested different permutations of the signs of $f_2$ and $f_3$ parameters. 

\begin{figure}[h]
\centering
  \includegraphics[width=\linewidth]{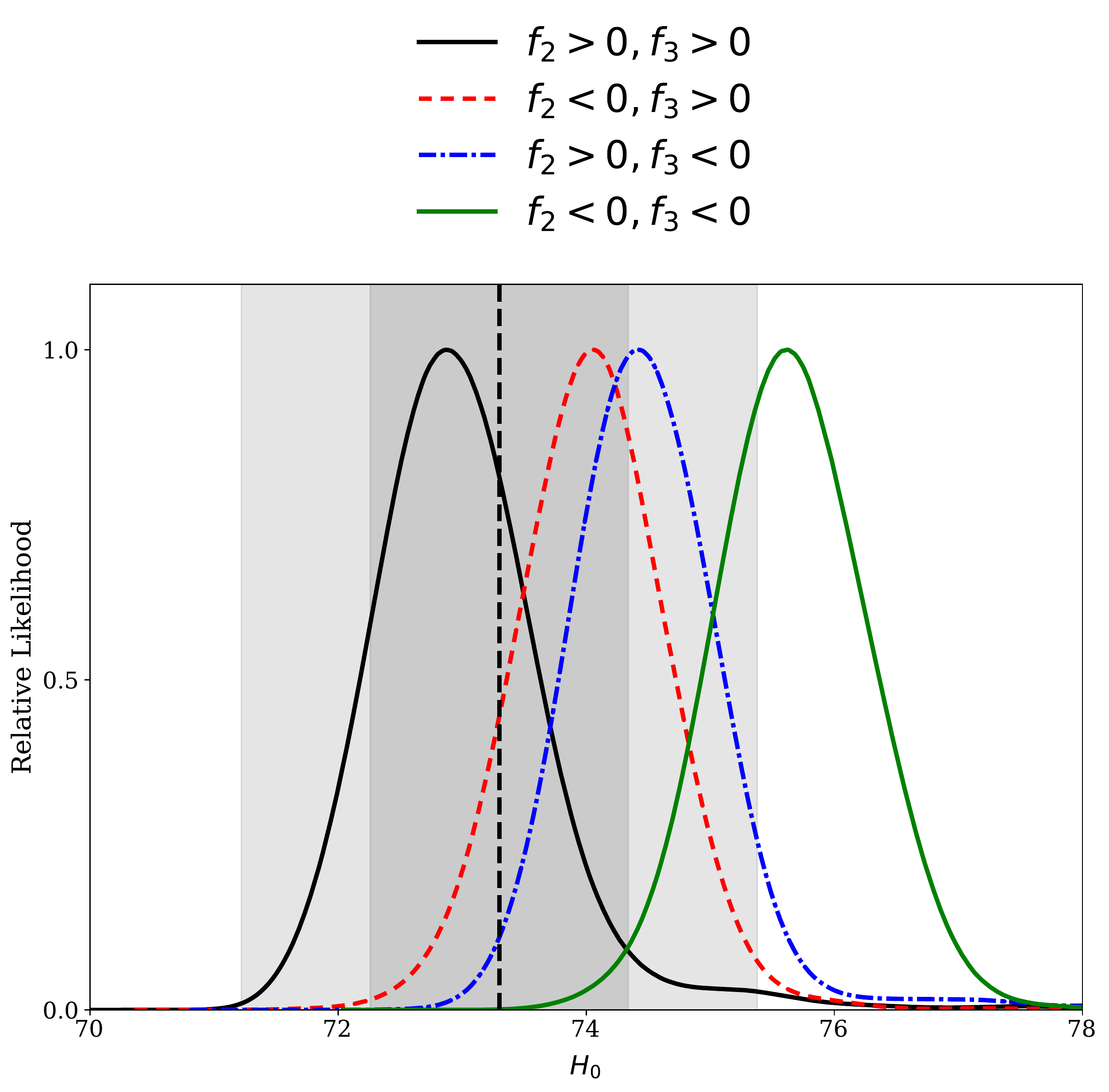}
  \caption{One-dimensional likelihoods for \textbf{Three-Step} models for $H_0$  based on ``CMB+BAO+SN'' data comparing the effects of  the signs of the 
 fraction of  dark energy in the second and third phases ($f_2$, $f_3$) with a fixed $f_1>0$.  }
  \label{fig:three-step-H0-likelihoods-dS-AdS}
\end{figure}

\begin{figure}[h]
\centering
  \includegraphics[width=\linewidth]{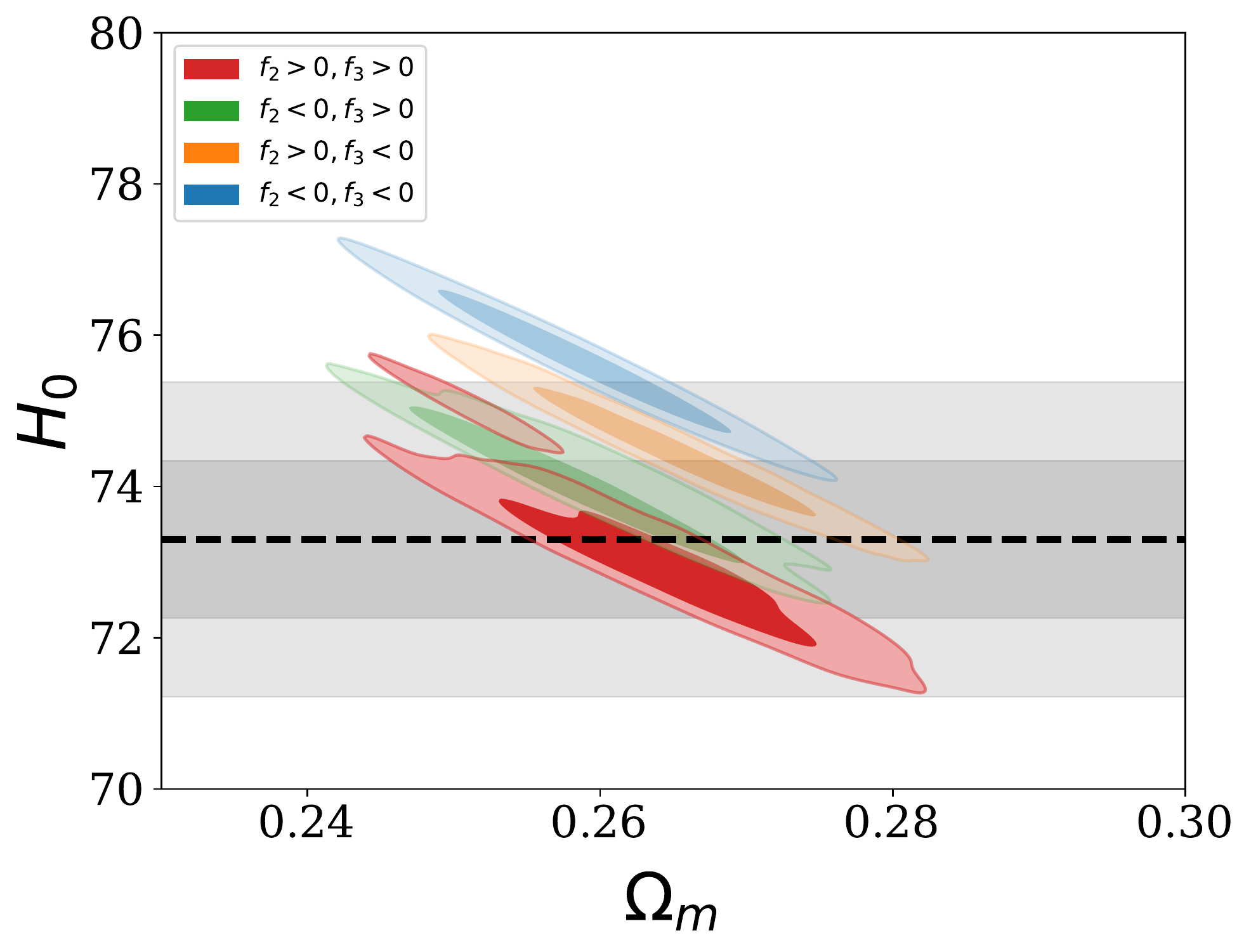}
 \caption{Contour plots for \textbf{Three-Step} models for $H_0$ vs. $\Omega_m$ based on ``CMB+BAO+SN'' data comparing the effects of the signs of the 
 fraction of  dark energy in the second and third phases ($f_2$, $f_3$) with a fixed $f_1>0$. Note that the shaded area shows measurement of $H_0$ done by SH0ES team and its $1\sigma$ error \citep{Riess:2021jrx}.}
  \label{fig:three-step-H0-omegam-contours-dS-AdS}
\end{figure}

According to Table.~\ref{tab:three-step-results-1} and Table.~\ref{tab:H0-tension-criteria} different cases of three-step class of models show consistency with SH0ES results. However, as we see from table \ref{tab:H0-tension-criteria}, 
Model 12 and Model 13 show the least tension with SH0ES measurements. In Model 12 all three steps are in dS  phase but in Model 13 the second step is chosen to be AdS phase. The corresponding results are presented in Fig.~\ref{fig:three-step-H0-likelihoods-dS-AdS} for the likelihood probabilities and also for contour plots of $H_0$ vs. $\Omega_m$ in Fig.~\ref{fig:three-step-H0-omegam-contours-dS-AdS}. A general conclusion is that having the second and third phases both to be in AdS type, $f_2, f_3 <0$, leads to higher values of $H_0$ compared to the cases where they are both in dS phases, $f_2, f_3 >0$, compare Models 12 and 15. However, the competition is more non-trivial when a dS phase and an AdS phase are both present, i.e. the cases where $f_2>0, f_3<0$ compared to the case where 
$f_2<0, f_3>0$, see Models 13 and 14. 

\begin{table*}[ht]
\centering
\begin{tabular} {c||c c c c c}
Parameter & Priors  & Priors  & Priors & Priors  \\
& Model 12 & Model 13 & Model 14 & Model 15 \\
\hline
\hline
{$1+z_{{\rm c},1}$} & 2500  & 2500  & 2500  & 2500  \\
\hline
{$f_1$} & 0.25  & 0.25  & 0.25   & 0.25  \\
\hline
{$w_1$} & $[0.5, 1]$  & $[0.5, 1]$ & $[0.5, 1]$ & $[0.5, 1]$  \\
\hline
{$n_1$} & $[-1, 1.5]$ & $[-1, 1.5]$ & $[-1, 1.5]$  & $[-1, 1.5]$ \\
\hline
{$1+z_{{\rm c},2}$} & 833  & 833 & 833 & 833 \\
\hline
{\boldmath$f_2$} & {\boldmath $+0.15$}   & {\boldmath $-0.15$}  & {\boldmath $+0.15$}  & {\boldmath $-0.15$}  \\
\hline
{$w_2$} & $[0.5, 1]$  & $[0.5, 1]$ & $[0.5, 1]$ & $[0.5, 1]$  \\
\hline
{$n_2$} & $ [-1, 1.5]$ & $[-1, 1.5]$ & $[-1, 1.5]$  & $[-1, 1.5]$  \\
\hline
{$1+z_{{\rm c},3}$} & 83.3 & 83.3 & 83.3 & 83.3  \\
\hline
{\boldmath$f_3$} & {\boldmath $+0.15$}   & {\boldmath $+0.15$}  & {\boldmath $-0.15$}  & {\boldmath $-0.15$}   \\
\hline
{$w_3$} & $[0.5, 1]$  & $[0.5, 1]$ & $[0.5, 1]$ & $[0.5, 1]$  \\
\hline
{$n_3$} & $[-1, 1.5]$ & $[-1, 1.5]$ & $[-1, 1.5]$  & $[-1, 1.5]$ \\
\end{tabular}
\caption{Priors for \textbf{Three-Step} models with different signs of the fractions of dark energy density $f_2, f_3$ in second and third steps.  We fix the first step in dS phase $f_1>0$ and test different permutations of dS or AdS phases for second and third steps. }
\label{tab:three-step-priors-1}
\end{table*}

\begin{table*}[ht]
\centering
\begin{tabular} {c||c c c c c}
Parameter & Best-fit & Best-fit & Best-fit & Best-fit \\
& Model 12 & Model 13 & Model 14 & Model 15 \\
\hline
\hline
{\boldmath$n_1$} & $ > 1.46$  & $> 1.46$ & $ > 1.46$  & $> 1.46$\\
\hline
{\boldmath$w_1 $} & $ > 0.991$ & $   0.543^{+0.014}_{-0.025}$ & $ > 0.994$ & $0.569^{+0.014}_{-0.026}$  \\
\hline
{\boldmath$n_2$} & $ > 0.600$  & $> 1.07$ & $> 1.06$ & $> 1.14$\\
\hline
{\boldmath$w_2$} & $ > 0.905$ & $> 0.865$ & $0.78^{+0.16}_{-0.11}$ & $> 0.940$ \\
\hline
{\boldmath$n_3$} & $ ---$  & $< -0.762$ & $> 0.682$  & $> 0.992$\\
\hline
{\boldmath$w_3$} & $ < 0.530$ & $< 0.521$ & $ > 0.865$  & $> 0.873$\\
\hline
{\boldmath$\Omega_m $} & $0.2634^{+0.0080}_{-0.0061}$  & $0.2586\pm 0.0073$ &$0.2641^{+0.0074}_{-0.0056}$   & $0.2585\pm 0.0076$ \\
\hline

{\boldmath$H_0$} & $73.09^{+0.46}_{-0.83}$ & $74.08\pm 0.75$ & $ 74.60^{+0.42}_{-0.73}$  & $75.70^{+0.56}_{-0.69}$\\
\hline
{\boldmath$S_8 $} & $0.861^{+0.019}_{-0.016}$ & $ 0.872\pm 0.019$ & $0.893^{+0.018}_{-0.015}$ & $0.903^{+0.018}_{-0.016}$ \\
\hline
{\boldmath$10^9 A_s $} & $ 1.979^{+0.028}_{-0.031}$ & $2.022\pm 0.032$ & $ 2.013^{+0.025}_{-0.030}$  & $2.057\pm 0.028$\\
\hline

{\boldmath$n_s $} & $ 1.0129\pm 0.0060$ & $1.0390^{+0.0053}_{-0.0039}$  & $0.9979^{+0.0039}_{-0.0046}$ & $1.0255^{+0.0044}_{-0.0039}$\\
\hline
{\boldmath$\tau $} & $ 0.0420^{+0.0059}_{-0.0069}$ & $0.0408\pm 0.0066$ & $ 0.0415^{+0.0057}_{-0.0066}$  & $0.0404\pm 0.0064$\\
\hline
\hline

\end{tabular}
	\caption{\label{tab:three-step-results-1} Constraints of $\%$68 limits and best-fit values for parameters of \textbf{Three-Step} models based on CMB+BAO+SN data (cf. table \ref{tab:three-step-priors-1}). }
\end{table*}

\section{Summary and Discussions}
\label{sec:Discussion}

We studied a phenomenological model in which  dark energy undergoes multiple transient stages.  At early time prior to transition, the dark energy 
behaves  like a small  cosmological constant term. At the transition time, dark energy comprises a noticeable fraction of the total energy density and falls off rapidly afterwards. The equation of state of the fluid after transition is represented by the parameter $w$ with the requirement $\frac{1}{3} < w\leq 1$ in order not to  modify the expansion history of universe drastically. 
While this is a phenomenological proposal which can mimic EDE scenario, but it may be realized theoretically as well. For example, this setup may be realized within the context of vacuum zero point energy of quantum fields in connection to the cosmological constant problem. Alternatively, this proposal may emerge from theories beyond SM of particle physics where the energy density of hidden sector do not interact with the SM fields while they contribute in the expansion history of the universe.

We have studied various  cases of single transition, double transition and triple transition
in dark energy density.  In the latter two cases we also allowed that the second and/or third components of 
dark energy to be either dS-like ($\rho_{_X} >0$) or AdS-like ($\rho_{_X} <0$).  As in standard EDE setup, having a larger value of $f_1$ yields to a larger value of $H_0$. In addition, AdS-like dark energy  yields to larger values of $H_0$. To solve the $H_0$ tension, as in EDE scenario, the first transition  is located sometime between the time of matter radiation equality and the surface of last scattering, say 
at the  redshift $z_{{\rm c},1} \sim 2500$. However, the second or third transitions can take place anytime after the CMB decoupling. We have considered the cases where these happen say at redshifts $z_{{\rm c},2}\sim 800$ and $z_{{\rm c},3}\sim 80$. Our investigations show that the resulting values of $H_0$ is not sensitive to the locations of the second or third transitions ($z_{{\rm c},2}$ and $z_{{\rm c},3}$) but it is largely sensitive to the value and the signs of the fraction of dark energy, $f_2$ and $f_3$.  Our analysis also shows that  to obtain  values  of $H_0$ comparable to the  value obtained from the 
local measurements requires that $n_s$ to  move towards  the Harrison-Zeldovich
scale invariant value. In all the examples which we studied so far the least tension in $H_0$ value occurs in a Three-Step model in which all of its phases of evolution are dS-like ( $f_{i}>0, i=1, 2, 3$). 

As we mentioned previously, in the current analysis we have concentrated only on the background evolution as the physics behind the dark energy transition was already complicated at the background level. A fully consistent analysis of the effects of transient dark energies requires the perturbations to be included as well. While this is beyond the scope of the current analysis but it is an interesting question to extend the current analysis where the perturbations are included as well.

\acknowledgments

We are grateful to Sunny Vagnozzi and E. Di Valentino
for insightful  comments and discussions. H. F. and A. T. would like to thank the ``Saramadan'' federation of Iran for the partial supports.  H.F. would like to thank YITP, Kyoto University for the hospitality during the workshop ``Gravity: Current challenge in black hole physics and cosmology'' where this work was in progress.

\begin{table*}[ht]
\centering
\resizebox{\textwidth}{!}{
\begin{tabular}{c c||c c||c c}
\hline
\hline
Model  & $H_0$ & Gaussian Tension  \\ \hline

$\Lambda$CDM&  $67.70 \pm 0.52$&$5 \sigma$  \\
\hline
\hline
One-Step cases (Tables: \ref{tab:onestep-priors-step-position} \& \ref{tab:one-step-priors-f-strength}) & &    \\
\hline
\hline
Model 1  & $66.43 \pm 0.54$&$7.9 \sigma $\\
Model 2 & $67.80 \pm 0.52$&$3.87 \sigma$ \\
Model 3 &$71.83^{+0.59}_{-0.67}$& $1.2 \sigma$ \\
Model 4 &$70.37 \pm 0.68$ &$2.3 \sigma $ \\
Model 5 & $73.84^{+0.49}_{-0.92}$ &$0.47 \sigma$ \\
\hline
\hline
Two-Step cases (Tables: \ref{tab:two-step-priors-second-step-position} , \ref{tab:two-step-priors-dS-AdS} \& \ref{tab:twostep-priors-3})  & &     \\
\hline
\hline
Model 6 & $71.92^{+0.39}_{-0.74}$&$2.46 \sigma $  \\
Model 7 & $71.55^{+0.52}_{-0.77}$&$2.83 \sigma $  \\
Model 8 & $70.11^{+0.59}_{-0.76}$&$2.66 \sigma $ \\
Model 9 & $72.19^{+0.54}_{-0.72}$&$0.94 \sigma $ \\
Model 10 & $73.10^{+0.50}_{-0.73}$&$0.17 \sigma $ \\
Model 11 & $74.70^{+0.34}_{-1.2}$&$1.27 \sigma $  \\
\hline
\hline
Three-Step cases (Table: \ref{tab:three-step-priors-1}) & &     \\
\hline
\hline
Model 12 & $73.09^{+0.46}_{-0.83}$&$0.07 \sigma $  \\
Model 13 & $74.08\pm 0.75$ & $0.58 \sigma $  \\
Model 14 & $74.60^{+0.42}_{-0.73}$&$1.02 \sigma $ \\
Model 15 & $75.70^{+0.56}_{-0.69}$&$1.76 \sigma $ \\

\hline
\hline

\end{tabular}}
\caption{$H_0$ tension criteria for One-Step, Two-Step and Three-Step models in comparison to $\Lambda$CDM model.}
\label{tab:H0-tension-criteria}
\end{table*}


\newpage
\newpage
\bibliography{references}

\end{document}